\documentclass[reprint,aps,prx,showpacs,showkeys,groupedaddress,longbibliography]{revtex4-1}

\usepackage{graphicx}        
\usepackage{grffile}         
\usepackage{graphics}
\usepackage{epsfig}
\usepackage{verbatim}
\usepackage{dcolumn}        
\usepackage{bm}            
\usepackage{amsmath}
\usepackage{mathtools}
\usepackage{physics}
\usepackage{hyperref}        
\hypersetup{colorlinks,breaklinks,
            linkcolor=black,urlcolor=black,
            anchorcolor=black,citecolor=black}
\usepackage{color}
\usepackage{float}
\newcommand{\EQ}[1]{Eq.~(\ref{eq:#1})}
\newcommand{\EQS}[2]{Eqs.~(\ref{eq:#1}) and (\ref{eq:#2})}
\newcommand{\FIG}[1]{Fig.~\ref{fig:#1}}

\def\bal#1\eal{\begin{align}#1\end{align}}
\newcommand{\be}{\begin{equation}}
\newcommand{\ee}{\end{equation}}
\newcommand{\ba}[1]{\begin{array}{*{#1}{c}}}
\newcommand{\ea}{\end{array}}

\newcommand{\OR}{\color{black}}

\definecolor{orange}{rgb}{1.0,0.4,0.2}
\newcommand{\MV}{\color{black}}
\newcommand{\old}{\color{black}} 

\graphicspath{{./Figures/}} 

\begin{document}
\title{The Mpemba index and anomalous relaxation}

\author{Israel Klich$^{1}$, Oren Raz$^{2}$, Ori Hirschberg$^{3}$ and Marija Vucelja$^{1}$}
\affiliation{$^{1}$ Department of Physics, University of Virginia, Charlottesville, VA 22904, USA\\ $^{2}
$ Department of Physics of Complex System, Weizmann Institute of Science, 76100 Rehovot, Israel, 
\\
$^{3}$ Courant Institute of Mathematical Sciences, New York University, New York, NY 10012, USA
}
\email{mvucelja@virginia.edu}
\email{orenraz@gmail.com}

\begin{abstract}
The Mpemba effect is a counter-intuitive relaxation phenomenon, where a system prepared at a hot temperature cools down faster than an identical system initiated at a cold temperature when both are quenched to an even colder bath. Such non-monotonic relaxations were observed in various systems, including water, magnetic alloys, polymers and driven granular gases. We analyze the Mpemba effect in Markovian dynamics and discover that   a stronger version of the effect often exists for a carefully chosen set of initial temperatures. In this \emph{strong Mpemba effect}, the relaxation time jumps to a smaller value leading to exponentially faster equilibration dynamics. The number of such special initial temperatures defines the \emph{Mpemba index}, whose parity is a topological property of the system. To demonstrate these concepts, we first analyze the different types of Mpemba relaxations in the mean field anti-ferromagnet Ising model, which demonstrates a surprisingly rich Mpemba phase diagram. Moreover, we show that the strong effect survives the thermodynamic limit, and that it is tightly connected with thermal overshoot -- in the relaxation process, the temperature of the relaxing system can decay non-monotonically as a function of time. Using the parity of the Mpemba index, we then study the occurrence of the strong Mpemba effect in a large class of thermal quench processes and show that it happens with non-zero probability even in the thermodynamic limit.  This is done by  introducing the \emph{isotropic} model for which we obtain analytical lower bound estimates for the probability of the strong Mpemba effects. Consequently, we expect that such exponentially faster relaxations can be observed experimentally in a wide variety of systems.
\end{abstract}

\keywords{Relaxation toward equilibrium, REM, Mpemba effect, Quench dynamics, Anomalous heating, Anomalous cooling, Antiferromagnet}

\maketitle

\section{Introduction}

The physics of thermal relaxation is rich with fascinating and often surprising behaviors. A particularly striking and counter-intuitive example is provided by the Mpemba effect. Known already to Aristotle~\cite{Aristotle} but named after a high-school student E. B. Mpemba~\cite{Mpemba}, the effect is commonly described as a curious phenomenon where initially prepared hot water freezes faster than cold water under otherwise identical macroscopic conditions, when both are cooled by the same cold bath. Due to the complexity of the phenomenon, the precise mechanism and conditions for the occurrence of the Mpemba effect have been under debate. Several explanations have been put forward to the particular mechanism for the Mpemba effect in water, highlighting possible roles for evaporation~\cite{mirabedin2017numerical}, supercooling~\cite{auerbach1995supercooling}, convection~\cite{Vynnycky2015243}, particular properties of the hydrogen bonds \cite{HydroBond_zhang2014hydrogen,DFT_Numerical_Mpemba}, freezing-point depression by solutes~\cite{Katz} and a difference in the nucleation temperatures of ice nucleation sites between samples~\cite{brownridge2011does}. Moreover, the status of the Mpemba effect in water as an experimental finding has been recently called into question~\cite{burridge2016questioning,katz2017reply}. Indeed, subtleties of the liquid-solid transition make the precise definition of the effect difficult. For example, when does freezing occur? How well is the freezing point defined? Is a small left-over amount of vapor or liquid tolerated in the description?  

It is possible to view the effect as a particular example of a relaxation process far from equilibrium: the Mpemba effect is defined by a \emph{quenching process} -- cooling through a quick change in the ambient temperature, achieved by putting the system in contact with a new, colder, thermal bath.  In contrast to quasi-static cooling, where the system is in equilibrium at each instant of the cooling process, quenching is, in general, a far-from-equilibrium process. Indeed, anomalous thermal relaxations are not unique to water, and similar effects have been observed in various other systems, e.g. magnetic alloys \cite{Magnetic_Mpemba}, carbon nanotube resonators \cite{Theory_Carbon_nano_greaney2011mpemba}, granular gases \cite{Granular_Mpemba_PhysRevLett}, clathrate hydrates \cite{paper:hydrates}, polymers \cite{18Mpembapoly} and even dilute atomic gas in an optical resonator \cite{keller2018quenches}.

Microscopically, the Mpemba effect occurs when the initially hotter system takes a non-equilibrium ``shortcut" in the system's state space and thus approaches the new equilibrium faster than the initially colder system. A phenomenological description of such behavior was recently proposed by Lu and Raz within the framework of Markovian dynamics~\cite{2016LuRaz}. In this picture, a Mpemba like behavior can be studied in a large variety of systems (as many processes in physics and chemistry are Markovian~\cite{van_Kampen}), and in particular, in small systems that can not be adequately described by macroscopic thermodynamics alone. An inverse Mpemba effect (associated with heating processes) can be described similarly. The suggested mechanism for the Mpemba effect raises several natural and important questions: (i) Does the mechanism require fine tuning of parameters, i.e., does it only occur in singular points of the model's parameter space? Is it robust to small changes in the system parameters?  
(ii) Does this mechanism survive the thermodynamic limit, or is it only a peculiarity of few-body systems such as those studied in~\cite{2016LuRaz}? One might intuitively expect that this mechanism  does not apply to macroscopic systems, since in such systems the probability distribution is highly concentrated on specific points of the system phase space -- those that minimize the free energy in equilibrium systems, and hence even if such ``shortcuts'' exist, the system cannot explore them.

The current manuscript contains the following contributions. First, using geometric insights on the relaxation dynamics in probability space we show that the Mpemba effect may be substantially enhanced on a discrete set of initial temperatures -- a phenomenon we call the \emph{strong Mpemba effect}. We show that these special initial temperatures can be  classified by an integer, which we name the \emph{Mpemba index}, and whose parity is a topological property of the system. Thus, the existence of a strong Mpemba effect is robust to small perturbations in the model parameters. Next, we study the effect in a thermodynamic system, focusing on a paradigmatic model: the mean-field Ising anti-ferromagnet, where a rich Mpemba-phase diagram is found. Using this model we demonstrate that even though in the thermodynamic limit the probability distribution is concentrated on specific points of phase space, the strong Mpemba effect still exists.  Interestingly, we show that the strong Mpemba effect is tightly connected with another type of anomalous thermal relaxation -- \emph{thermal overshooting}, in which the temperature of the system relaxes non-monotonically in time and overshoots the bath's temperature. We then provide an exact analytical calculation of strong Mpemba effect probability for an arbitrarily chosen set of energy levels in an idealized ``isotropic'' model. Comparison of these analytic results with the dynamics of the same set of energy levels with random barriers, gives a surprisingly good quantitative agreement. Lastly, we numerically study the strong Mpemba effect in the random energy model (REM) with random barriers, and find the scaling of the probability of the strong effect with the system size. This scaling suggests that again the effect can be observed in the thermodynamic limit.

The manuscript is organized as follows. In section~\ref{sec:setup} we give details on the explicit form of the Markovian dynamics, and in section~\ref{sec:strong} we define the strong Mpemba effect and describe its geometric meaning.  
Next, in section~\ref{se:Ising_model} a study of the dynamics of an anti-ferromagnetic Ising model on a complete bipartite graph reveals a remarkable phase diagram exhibiting a variety of phases with various values of the Mpemba index, ${\mathcal I}_M=0,1,2$ and phases with both direct and inverse strong Mpemba effect. In section \ref{Sec:Mpemba_in_Thermodynamic_limit} we show that the Mpemba effect also appears in the thermodynamic limit of the anti-ferromagnetic Ising model, and lastly, in section~\ref{Sec:Overshoot} we show that in this specific model the strong Mpemba effect implies overshoot in the temperature during relaxation. In section~\ref{subse:Isotropic_ensemble} we study the probability of the occurrence of an odd Mpemba index for a system with random barriers taken from a very wide distribution and a fixed set of energies. For this purpose, we use as a model a statistically isotropic ensemble of second eigenvectors of the driving rate matrix. In particular, we find that the probability of a strong Mpemba effect is inversely proportional to the bath temperature, where the proportionality constant depends on the first few moments of the energy level distribution. In section~\ref{subse:REM0} we study the Mpemba effect in REM with random barriers. 

\section{\label{sec:setup} Setup and Definitions}

We consider Markovian relaxation dynamics, as given by the Master equation \footnote{For simplicity, we consider here only ergodic finite state systems. Much of the analysis can be easily generalized to infinite systems as well.}
\bal 
\label{eq:MasterEq}
\partial _t \bm p= R \bm p, 
\eal 
where $\bm p=(p_{1},p_{2},...)$, and $p_i(t)$ is the probability to be in state $i$ at time $t$. Here, the  off-diagonal matrix element $R_{ij}$ is the rate (probability per unit time) to jump from state $j$ to $i$. The state $i$ of the system is associated with an energy $E_i$, and we focus on relaxation dynamics for which the steady state of \EQ{MasterEq}) is given by the Boltzmann distribution, 
\bal
{\pi _i(T_b)}\equiv \frac{e^{-\beta_b E_i}}{Z(T_b)}, 
\eal 
where $T_b$ is the temperature of the bath, $Z(T_b)=\sum_i e^{-\beta_b E_i}$ is the partition function at $T_b$, and throughout the paper  $\beta \equiv (k_BT)^{-1}$ is the inverse temperature (in particular $\beta_b = (k_BT_b)^{-1}$). Moreover, we also assume that the rate matrix $R$  obeys detailed balance,
\bal
R_{ij} e^{-\beta_b E_j} = R_{ji} e^{-\beta_b E_i},
\eal 
and thus can be written in the form (see e.g. \cite{2011MandalJarzynski}):
\bal
R_{ij} = \begin{cases}
\Gamma e^{-\beta_b (B_{ij} - E_j)},& i\neq j\\
-\sum _{k\neq j}R_{{kj}}, & i=j\\
\end{cases}\label{eq:Driving},
\eal
where  $B_{ij}=B_{ji}$ can be interpreted as the barrier between the states, and $\Gamma$ is a constant with the proper units.  At long times, the Markov matrix (\ref{eq:Driving}) drives an arbitrary initial distribution to the Boltzmann distribution associated with the bath temperature $T_b$. 
Note that if the matrix $R$ does not satisfy the detailed balance condition, its steady state does not represent an equilibrium since it has non-vanishing current cycles. Interestingly, a direct and an inverse Mpemba-like effects were recently discovered in driven granular gases where detailed balance is violated \cite{Granular_Mpemba_PhysRevLett}. Although our approach may be useful also for such non-equilibrium steady states, for simplicity we limit our discussion to systems obeying detailed balance. 

In the Mpemba effect scenario, the initial condition for \EQ{MasterEq} is the thermal equilibrium for some temperature $T\neq T_{b}$,
\bal
p_i(T;t=0)={\pi _i(T)}\equiv \frac{e^{-\beta E_i}}{Z(T)}.
\eal
During the relaxation process, the distribution $\bm p$ --- i.e. the solution of \EQ{MasterEq} -- can be written as
\bal
\label{eq:ProbExpansion}
\bm p(T;t) = e^{Rt}\bm \pi(T)= \bm \pi(T_b) + \sum _{i > 1} a_i(T) e^{\lambda_{i}t} \bm v_i,
\eal 
where the rate matrix $R$ has (right) eigenvectors $\bm v_i$ and eigenvalues $\lambda_i$,
\bal
\label{eq:Def_v}
R\bm v_i = \lambda _i \bm v_i.
\eal 
The largest eigenvalue of $R$, $\lambda _1 = 0$, is associated with the stationary (equilibrium) distribution $\bm \pi(T_b)$, whereas all the other eigenvalues have negative real part, $0 > \Re \lambda_2 \geq \Re \lambda_3 \geq ...$, and they correspond to the relaxation rates of the system. The equilibration timescale is typically characterized by   $-(\Re\lambda_{2})^{-1}$ \footnote{For detailed balance matrices $R$, the eigenvalues are in fact all real. We use this more general notation as our discussion can also be relevant to $R$'s that do not satisfy detailed balance.}.

Any detailed balance  matrix $R$ can be brought to a symmetric form $\tilde{R}$ via the similarity transformation,
\bal
\tilde{R} = F^{1/2} R F^{-1/2},\label{eq:SymDriving}
\eal
where $F_{ij} \equiv e^{\beta_b E_j}\delta_{ij}$.
The matrix $\tilde{R}$ has the same eigenvalues as $R$, and has an orthogonal set of real eigenvectors. In particular $\bm f_i \equiv F^{1/2}\bm v_i$ are eigenvectors of $\tilde{R}$ with eigenvalues $\lambda_{i}$. The $\bm f_{i}$ form an orthogonal basis, with $\bm f_i\cdot \bm f_j= \left(\bm v_{i}\cdot F \bm v_{j}\right)\delta_{ij}$. This form will be useful in what follows.

\subsection*{The Mpemba effect}
A simple criterion for the presence of a Mpemba effect for the relaxation process in Eq. \eqref{eq:MasterEq} was given by Lu and Raz \cite{2016LuRaz}. When $|\Re\lambda_2| < |\Re\lambda_3|$ (namely when they are not equal), the probability distribution (\ref{eq:ProbExpansion}) can be approximated, after a long time, as 
\bal 
\label{eq:p_long_time_limit}
\bm p(T;t)\approx \bm \pi(T_b) + a_2(T)e^{\lambda_2t}\bm v_2.
\eal 
In this case the Mpemba effect is characterized by the existence of three temperatures: hot, cold and the bath, ($T_h> T_c> T_b$, respectively), such that \footnote{The above definition to the Mpemba effect is readily generalizable for the degenerate case, $|\Re\lambda_2| = |\Re\lambda_3|$.}
\bal
|a_2(T_h)| < |a_2(T_c)|.
\eal
The coefficient $a_2$ can be derived as follows:  multiplying \EQ{ProbExpansion} with $\bm f_2 F^{1/2}$ from the left, substituting ${\bm v}_i =F^{-1/2}{\bm f}_i $ and using the fact that $\bm f_i$ form an orthogonal basis, one get  $\bm f_2 \cdot F^{1/2}\bm p(T; t) = a_2 (T) ||\bm f_2||^2 e^{\lambda _2 t}$. Therefore for an evolution starting at a given initial probability ${\bm p}_{init}$ we have that $a_2$ is the corresponding overlap coefficient between the initial probability and the second eigenvector $\bm f_2$:
\bal 
\label{eq:a2}
a_2 = \frac{\bm f_2 \cdot F^{1/2}\bm p_{init}}{||\bm f_2||^2}. 
\eal 
At the bath temperature this coefficient vanishes, $a_2(T_b) = 0$ (as in this case $F^{1/2}\bm p_{init}= {\bm f_1}$, which is orthogonal to $\bm f_2$), and it increases in absolute value as the initial temperature departs from the bath. Therefore, to determine whether the Mpemba effect exists one has to look for non-monotonicity of $a_2(T)$. 

In the next section, we define the strong Mpemba effect, introduce an index to characterize the strong effect and describe the geometrical interpretation of the effect. 

\section{\label{sec:strong} The strong Mpemba effect, its index and its parity}
  
Our first contribution is the observation that  a stronger effect (even shorter relaxation time) can occur: a process where there exists a temperature $T_M\neq T_{b}$ such that 
\bal \label{eq:a2=0} a_{2}(T_M)=0.\eal 
We call such a situation a \emph{strong direct Mpemba effect} if $T_M>T_b$ and a \emph{strong inverse Mpemba effect} if $T_M<T_b$, as at $T_M$ the relaxation process is exponentially faster than for initial temperatures slightly below or above it. Since there is essentially no difference between the direct and inverse effects, we refer to both of them as \emph{strong Mpemba effects}.  The strong Mpemba effect implies the existence of the ``weak'' effect, as in order to cross zero, $a_2$ has to be a non-monotonic function of temperature (because $a_2(T_b) = 0$, whereas $a_2 \neq 0$ slightly above and below $T_b$) \footnote{Since the relaxation described by Eq. \eqref{eq:MasterEq} reaches the bath's Boltzmann distribution at infinite times, the actual observed relaxation time may depend on the choice of a distance function on the probability simplex. The distance function may be relative entropy or other measures, as long as the exponential ratio between the time dependent coefficients of ${v_3}$ and $v_2$ is not compensated by the fact that the distances are measured along different directions. In other words, the metric does not grow exponentially faster in one direction compared to the other. }.

To study the strong Mpemba effect, we define the Mpemba indices  as:
\bal
\mathcal{I}_M^{dir} \equiv \text{$\#$ of zeros of }a_2(T),\,\, T_b<T<\infty,\nonumber\\
\mathcal{I}_M^{inv} \equiv \text{$\#$ of zeros of }a_2(T),\,\, 0<T<T_b,
\eal 
and the total index as:
\bal
\mathcal{I}_M = \mathcal{I}_M^{dir} + \mathcal{I}_M^{inv}.
\eal 
\OR{$\mathcal{I}_M^{dir}$ changes its value when the number of zero crossing of the graph of $a_2(T)$ changes in the interval $T_b < T <\infty$. This implies that it is robust, as discussed in section~\ref{sec:robustness}. 

\subsection{The geometry of the strong Mpemba effect}

The geometry of the problem is schematically illustrated in \FIG{fig_01_TriangelFig1} for a three state system. The set of all points ${\bm p} = (p_1,p_2,p_3)$ which are normalized $(p_1+p_2+p_3=1)$ and non-negative ($p_i\geq0$) form the \emph{probability simplex} -- the blue triangle in the figure. The set of all Boltzmann distributions ${\bm \pi}(T)$ form the \emph{Boltzmann curve} -- the red line, which has two boundaries: the ${\bm \pi}(T=0)$ where the probability is concentrated at the lowest energy state (blue point), and ${\bm \pi}(T=\infty)$ -- the maximally mixed state in the middle of the simplex where all the states are equally probable (the red point). \MV The set of all ${\bm p}$ for which $a_2=0$ is illustrated by the intersection of the green plane with the blue triangle.\old The Boltzmann curve intersect the $a_2=0$ plane at  ${\bm \pi}({T_b})$  (pink point) since, being the equilibrium distribution at $T_b$,  $a_2(T_b)$ vanishes by definition. However, in the specific example, the two boundaries of the Boltzmann curve are both on the same side of the $a_2=0$ hyperplane, therefore there must be another point -- marked by the green point in the figure -- at which the Boltzmann line cross the $a_2=0$ hyperplane again. This point corresponds to $T_M$ where there is a strong Mpemba effect.  Topologically, having co-dimension 1, the $a_2=0$ hyperplane separates the probability simplex into two disjoint sets. The parity of the number of times a continuous curve crosses this hyperplane depends only on the two boundaries of the curve: if they are both in the same set, then the number of crossing is even -- and hence the parity of $\mathcal{I}_M$ is odd (as it does not count the crossing at $T_b$), and if they are in a different set then the number of crossing must be odd, with an even parity for $\mathcal{I}_M$.

\begin{figure}[h]
\includegraphics[width=\linewidth]{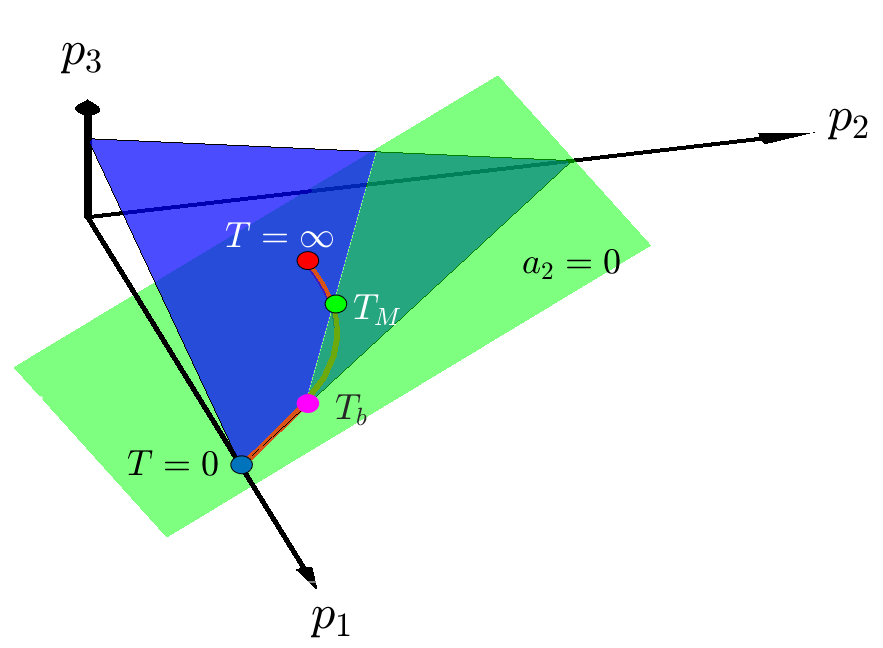}
\caption{\label{fig:fig_01_TriangelFig1}The geometry of the strong Mpemba effect in a three-state system.  The probability simplex is illustrated by the blue triangle. The set of all equilibrium distributions form the red curve. The blue and red points on the curve correspond to $T=0$ and $T=\infty$.  The Mpemba index is non-zero if the equilibrium curve crosses the $a_2=0$ plane (green) not only at  the bath temperature (illustrated here by the pink point) but also at some other temperature (the green point).}
\end{figure} 

Given $T_{b}$, a sufficient condition for the strong direct Mpemba effect to occur is obtained by determining whether $a_{2}$ changes sign going from $a_{2}(T_{b}+\varepsilon)$ to $a_{2}(T = \infty)$. This can be expressed by 
\bal
\label{eq:index_dir}
\mathcal{P} (\mathcal{I}_M^{dir})& \equiv \theta\left(-\partial_Ta_2(T)\vert_{T=T_{b}}a_2(T= \infty)\right),
\eal
where we used $a_{2}(T_{b}+\varepsilon)\approx \partial_Ta_2\vert_{T=T_{b}}\varepsilon$ and  $\theta$ is the Heaviside step function. The argument of the step function in \EQ{index_dir} is positive if $a_2(T)$ has an odd number of zero crossings, thus \EQ{index_dir} described the \emph{parity} of the number of zeros. In particular, if $\mathcal{P}(\mathcal{I}_M^{dir})\neq 0$, we are assured to have at least one crossing, and so $\mathcal{P}(\mathcal{I}_M^{dir})$ serves as a lower bound on the number of initial temperatures for which the direct strong Mpemba effect occurs. 
Similarly, the parity for the strong inverse Mpemba effect is  
\bal
\label{eq:index_inv}
\mathcal{P} (\mathcal{I}_M^{inv}) &
\equiv \theta\left(\partial_Ta_2(T)\vert_{T=T_{b}}a_2(T= 0)\right),
\eal
and the parity of the strong Mpemba effect is
\bal
\mathcal{P} (\mathcal{I}_M) &
\equiv \theta\left(a_2(T=\infty)a_2(T= 0)\right).
\eal

As already mentioned above, in some situations there are zeros of $a_2(T)$ that are not accompanied with a sign change. This  happens when $a_2(T_M)=0$ and $a_2'(T_M)=0$ simultaneously. Such points appear on the boundary between areas in parameter space with $\mathcal{I}_M=0$ and $\mathcal{I}_M=2$, e.g. on the line separating purple and green areas in \FIG{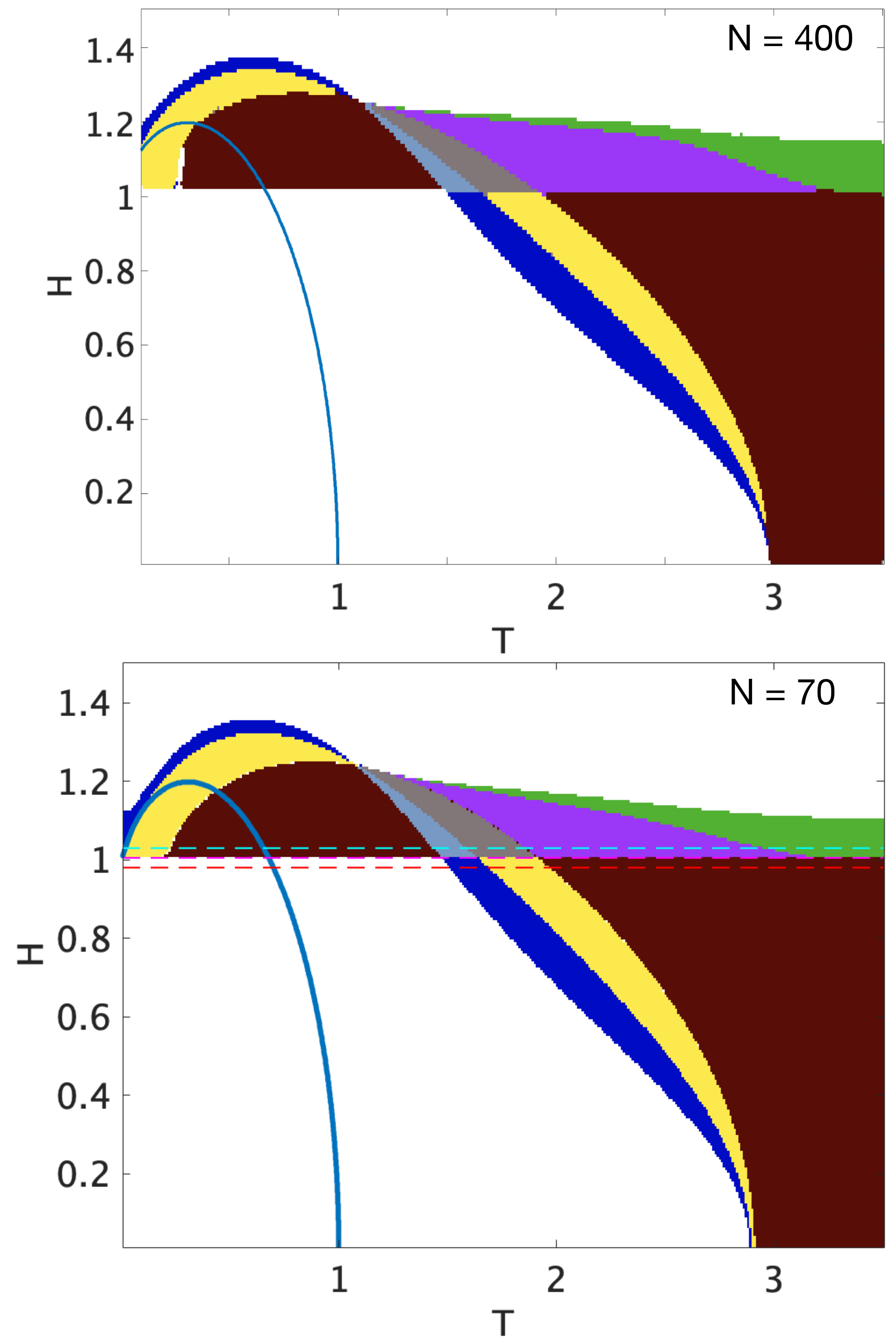}. In these cases, $\mathcal{P} (\mathcal{I}_M)$ is no longer the exact parity, but still serves as a lower bound to the number of crossings. 

\FIG{fig_01_TriangelFig1} provides a helpful three-dimensional picture for the strong Mpemba effect in a three-state system. In fact, the existence of the strong Mpemba effect is always essentially a three-dimensional problem, when projected to the proper sub-plane: as discussed above, the strong effect can be deduced from the relative directions of the following three vectors: (i) The tangent to the equilibrium line at the bath temperature; \MV{(ii) The vector connecting ${\bm\pi}(T_b)$ and ${\bm \pi}(T=\infty)$ (or ${\bm\pi}(T=0)$ for the inverse effect)\old; and (iii) the direction of the slowest dynamic, ${\bm v_2}$ defined in \EQ{Def_v}). This observation plays a crucial role in Sec.(\ref{subse:Isotropic_ensemble}), where we estimate the probability to observe the strong Mpemba effect in a class of random models. 

\MV
\subsection{Robustness of the Mpemba Index}
\label{sec:robustness}

The above geometric interpretation implies that the Mpemba index is a robust quantity, as we discuss next. Consider a small perturbation of order $\epsilon$ in the physical quantities, i.e. in the energies, the barriers and the temperature of the bath. The corresponding change in $R$ is also of order $\epsilon$. This perturbation in $R$ changes both ${\bm p}_{init}(T)$ and ${\bm f}_2$, and hence by Eq.(\ref{eq:a2}), also the graph of $a_2(T)$ in the relevant interval of $T$. But it does not change  ${\bm \pi}(T=\infty)$, which is always the maximally symmetric point. Similarly, ${\bm \pi}(T=0)$  is the lowest energy point, which changes only if the perturbation changes the ground state by an energy level crossing.

For the perturbation in the physical parameters to changes the number of zero crossings, one of the following cases has to occur: (i) ${\bm p}_{init}$ can change abruptly even with a small perturbation in $R$, if there is a first order phase transition in the system and therefore the equilibrium distribution changes discontinuously. An example for such a case is discussed in Sec. \ref{Sec:Mpemba_in_Thermodynamic_limit}.  Note that for this to happens, the perturbation in the relevant parameters has to be large enough compared to the distance from the value at which there is a first order phase transition. (ii) ${\bm f}_2$ can change abruptly when the perturbation changes the order of $\lambda_2$ and $\lambda_3$, namely the order of the eigenvalues changes and therefore the direction of the eigenvector jumps.  As the perturbations of the eigenvalues are of order $\epsilon$ too, the spectral gap  $\lambda_2- \lambda_3$ defines the stability region in which these changes are not expected. In other words, if the perturbation in $R$ is small compared to $\lambda_2- \lambda_3$ then such a jump is not expected. (iii) A small perturbation in $R$ can cause two zeros to ``annihilate'' each other in a saddle-node bifurcation, and similarly two new zeros can be generated. However, in these cases the parity of $\mathcal{I}_M$ does not change. (iv) A zero can move through $T_b$ (this is analogues to a transcritical bifurcation). In such a case both $\mathcal{I}_M^{dir}$ and $\mathcal{I}_M^{inv}$ change by one, but the parity of $\mathcal{I}_M$ does not change. Lastly, (v) a zero can ``vanish'' in the $T\rightarrow\infty$ limit or at $T=0$.  But as discussed above, the end points ${\bm\pi}(T=0)$ and ${\bm\pi}(T=\infty)$ do not move, so this can only happen if ${\bm f}_2$ changes its direction. 

From the five cases discussed above we can conclude that  the parity of $\mathcal{I}_M$ can change only if there is a phase transition in the system, $\lambda_3$ becomes larger than $\lambda_2$ or if the perturbation changes the sign of one of $a_2(0)$ and $a_2(\infty)$.	Therefore, the parity is stable in some range, which depends on details as the spectral gap $\lambda_2-\lambda_3$, the distance of parameters from a phase transition and the angle between ${\bm\pi}(T=0)$, ${\bm\pi}(T=\infty)$ and ${\bm f}_2$. 

The above argument for stability can be explained using the geometric picture for the Mpemba effect discussed above.  In any system, the $a_2=0$ hyperplane separates the probability space into two disjoint sets as discussed in the three state system above. The Boltzmann curve intersects the $a_2=0$ hyperplane at $T=T_b$. Any additional temperature $T_M$ for which $a_2(T_M)=0$ is an intersection between the curve and the hyperplane, and $\mathcal{I}_M$ counts these additional intersections. Topologically, the parity of the number of crossing between a continuous curve and a hyperplane of codimension 1 depends only on the boundaries of the equilibrium curve, namely in our case on ${\bm \pi}(T=0)$ and ${\bm \pi}(T=\infty)$. If they are both on the same side of the hyperplane then the number of crossing is even, and if they are on different sides then the number of crossing is odd.}

To appreciate the above topological aspect, let us contrast it with a (possibly only hypothetical) ``super strong Mpemba effect", where there exist a temperature $T_{SM}$ at which $a_2(T_{SM})=a_3(T_{SM})=a_4(T_{SM})=0$. In other words, consider a case where the coefficients of ${\bm \pi}(T_{SM})$ vanish along both ${\bm v}_2$,  ${\bm v}_3$ and ${\bm v}_4$. This implies an even faster relaxation than the  strong effect.  The condition for this super strong effect, $a_2(T_{SM})=a_3(T_{SM})=a_4(T_{SM})=0$, defines a hyperplane of co-dimension 3, that can intersect the probability simplex (which is of co-dimension 1) in a co-dimension 2 hyperplane. This hyperplane does not separate the space into two disjoint sets, and the topological argument does not work anymore. An example for this fact, consider a codimension 2 hyperplane in a 3D space, which is a straight line. It does not separate the 3D space into two disjoint sets as the a plane does. Now consider an equilibrium locus that crosses the super-strong straight line in a 3D probability simplex. A small perturbation in the model parameters deforms a bit the equilibrium locus, and generically separates the equilibrium locus from the straight line. Therefore, even an infinitesimal perturbation can (and usually does) destroy the super-strong Mpemba effect. In this sense, the strong Mpemba effect has a  topological stability,  but the super strong effect does not. We therefore do not expect to observe the super strong effect in systems which are not fine tuned. 
\old

\section{Mean Field Ising Antiferromagnetic model with Glauber dynamics}
\label{se:Ising_model}

The mechanism for the  Mpemba effect suggested in \cite{2016LuRaz} was so far demonstrated only in microscopic systems with a few degrees of freedom. However, all the experimental observations of similar effects are in macroscopic systems, with a huge number of micro-states.  To discuss the applicability of the mechanism in such macroscopic systems, we next consider the Ising model, with anti-ferromagnetic interactions and mean-field connectivity, on a complete bipartite graph. This is a classical many-body model which has been studied extensively, and whose phase diagram can be calculated exactly (see figure 6 in \cite{Mean_field_MJP_1997unified}). As described below, this model shows a rich Mpemba behavior, which survives the thermodynamic limit.

\subsection{The Model}
In mean-field models, each spin interacts equally with all the other spins in the system. To generate a model of antiferromagnetic interactions in the mean field approximation, we consider a system with a total number of $N$ spins, half of them on each ``sub-lattice'' or sub-graph. Each spin interacts antiferromagnetically with all the spins in the other sub-graph, but spins on the same sub-graph do not interact at all. The interaction strength between the spins is fixed. This type of interaction can lead to an ``antiferromagnetic phase'' in which the spins in one sub-lattice are predominantly in the up state, while most spins in the other sub-graph point down.   

Let $N_{1,\uparrow}$, $N_{1,\downarrow}$ ($N_{2,\uparrow}$, $N_{2,\downarrow}$) be the number of spins pointing up  and down on sub-graph $1$ (sub-graph $2$). We define the two magnetization densities on sub-graphs $1$ and $2$  as:
\bal
&x_1 \equiv \frac{N_{1,\uparrow} - N_{1,\downarrow}}{N/2}
\quad \text{and}\quad x_2 \equiv \frac{N_{2,\uparrow} - N_{2,\downarrow}}{N/2}.
\eal
Although the system has $2^N$ different microstates, all microstates that correspond to the same values of $N_{1,\uparrow}$ and $N_{2,\uparrow}$ are  equivalent, since the interaction strength is ``position'' independent (mean field). Thus, the Hamiltonian of this model is only a function of $x_1$ and $x_2$ and is given by
\bal
\label{eq:Antiferro_hamiltonian}
\mathcal{H} = \frac{N}{2}\left[-Jx_1x_2 - \mu H(x_1+x_2)\right],
\eal
where $J$ is the coupling constant, $H$ is the external magnetic field and $\mu$ is the magnetic moment. In the antiferromagnetic case, the coupling constant is negative, $J<0$, and for simplicity we choose the units such that $J=-1$ and $\mu=1$ . 

The dynamics we consider for this model is Glauber dynamics, with only a single spin flip transitions allowed. Under this assumption, the rates of flipping a spin up or down in sub-graphs $1$ and $2$ are given by 
\bal \label{eq:GlauberIsingRates}
 R^{u_1}(x_1,x_2) &= \frac{(1-x_1)/2}{1+e^{(-2x_2-2H)/T_b}},\nonumber
 \\
 R^{u_2}(x_1,x_2) &= \frac{(1-x_2)/2}{1+e^{(-2x_1-2H)/T_b}},
 \nonumber\\
 R^{d_1}(x_1,x_2) &= \frac{(1+x_1)/2}{1+e^{(2x_2+2H)/T_b}},
 \nonumber
 \\
 R^{d_2}(x_1,x_2) &=\frac{(1+x_2)/2}{1+e^{(2x_1+2H)/T_b}}.
 \eal
where $R^{u_1}(x_1,x_2)$ is the rate of flipping a spin up in sub-graph $1$ and $R^{d_2}(x_1,x_2)$ is the rate of flipping a spin down in sub-graph $2$.  The  numerators in Eqs.(\ref{eq:GlauberIsingRates}) are the combinatorial factors that take into account how many spins can be flipped in the specific state of the system, and the denominator is the standard Glauber factor, $1/(1+e^{\beta_b\Delta E})$, where $\Delta E$ is the difference of energies before and after the spin flip~\cite{Glauber}.  

\subsection{Mpemba Index phase diagram}

The Mpemba-index phase diagram of this model was calculated numerically for $N=400$, and is shown in the upper panel of \FIG{fig_02_phase_diagram_v01.pdf}. At each point in the figure, that is -- for each temperature $T_b$ and magnetic field $H$ of the environment, we  calculated (numerically) the coefficient $a_{2}(T)$ of the slowest relevant eigenvector of the corresponding Glauber dynamics (Eq. \ref{eq:GlauberIsingRates}) at each point along the equilibrium line. From the monotonicity and zero crossing of these coefficients, $a_{2}(T)$, we have deduced what types of Mpemba effect exist at this point. The phase diagram in Fig.(\ref{fig:fig_02_phase_diagram_v01.pdf}) is quite rich and has 8 different phases, differentiated through their colors,  including regions with odd and even Mpemba index, existing for the direct inverse or both effects. 

To make sure that the observed phase diagram is not dominated by the number of spins in the system,  we repeated this calculation with $N=70$  and checked that the phase diagram looks essentially the same (Fig. \ref{fig:fig_02_phase_diagram_v01.pdf}, lower panel).  Moreover, we have checked other form of rates -- Metropolis and heat-bath dynamics, both with single spin flips only. Although the exact locations of the different phases are not identical in the different dynamics, the main features in the phase diagram are similar in all of them. An example for such a feature is the line at $H=1$ across which the Mpemba phase changes. To explain this feature, we next consider the thermodynamic limit of this model. 
 
\begin{figure}[h]
\includegraphics[width=0.91\linewidth]{fig_02_phase_diagram_v01.pdf}
\caption{\label{fig:fig_02_phase_diagram_v01.pdf}The mean field anti-ferromagnetic Ising model Mpemba phase diagram. \textbf{Upper panel} -- the phase diagram calculated for $N=400$ spins. The are 8 different Mpemba-phases in this systems: (i) White  -- no direct nor inverse Mpemba effects; (ii) Blue -- Weak direct and no inverse effects; (iii) Green -- weak inverse and no direct; (iv) Burgundy -- strong inverse with $\mathcal{I}_M^{inv}=1$ and no direct; (v) Violet -- strong inverse with $\mathcal{I}_M^{inv}=2$ and no direct; (vi) Light Blue --  strong inverse with $\mathcal{I}_M^{inv}=1$ and weak direct; (vii) Gray --  strong direct with $\mathcal{I}_M^{dir}=1$ and strong inverse with $\mathcal{I}_M^{inv}=1$; and (viii) Yellow -- strong direct with $\mathcal{I}_M^{inv}=1$ and weak inverse. The blue line is the anti-ferromagnet to paramagnet phase transition line of this model, calculated in~\cite{Mean_field_MJP_1997unified}. \textbf{Lower panel} -- the same calculation for $N=70$ spins. Although the exact location of boundaries between the different phases is not identical in the two computations, the overall structure is the same. The three dashed lines are lines of constant magnetic field $H=1.03$, $H=1$ and $H=0.97$, and their corresponding equilibrium loci in the thermodynamic limit are given in~\FIG{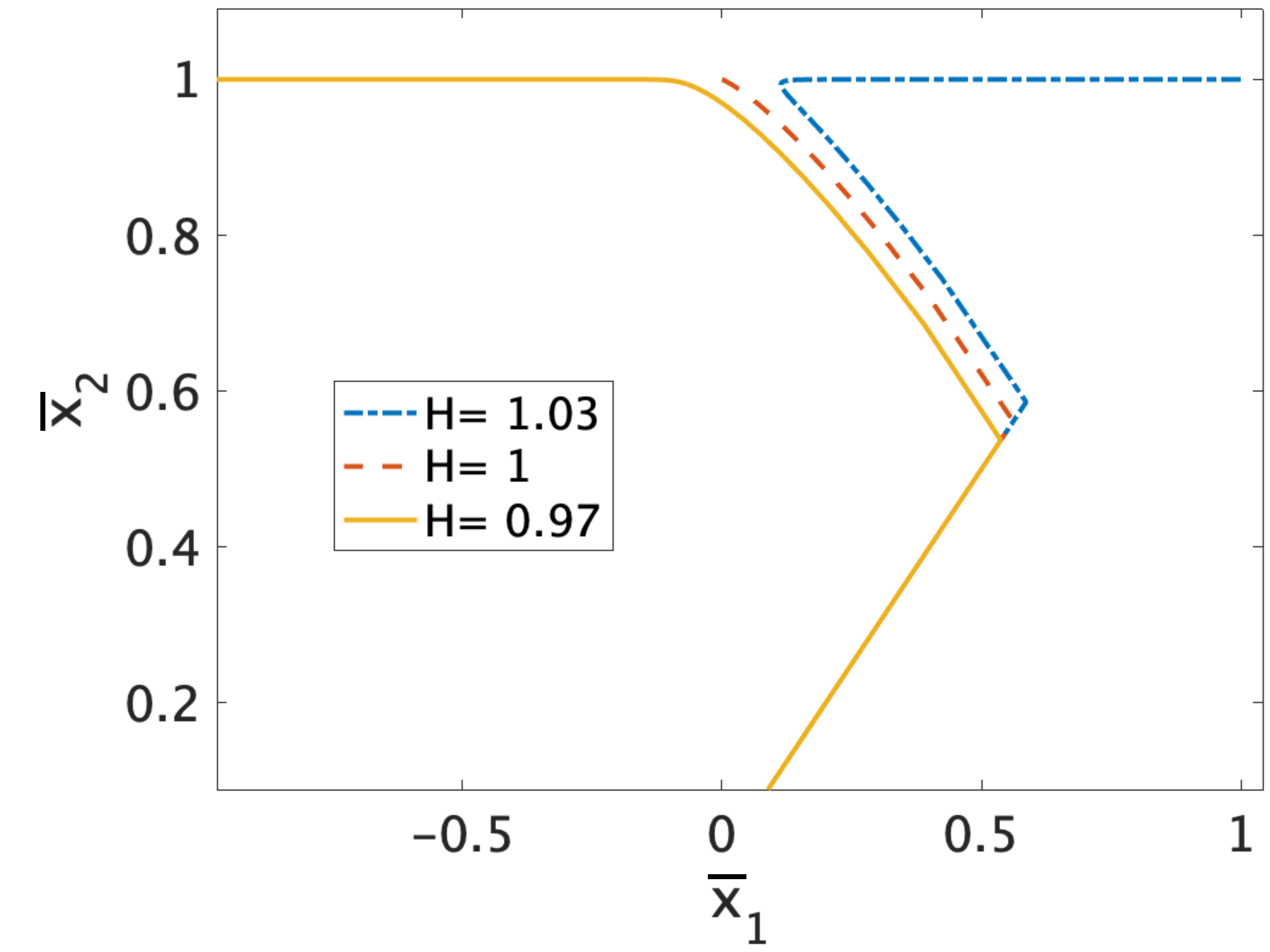}. Note that the phase diagram has an abrupt jump at $H=1$, which corresponds to the jump in the equilibrium locus in~\FIG{fig_03_Equilibrium_Line_Different_H_v01.pdf}.}
\end{figure} 

\subsection{The Thermodynamic  Limit}
Let  us take the thermodynamic ($N\rightarrow\infty$) limit for  the mean field anti-ferromagnet model described above. To this end, we first write explicitly the master equation using all the single flip rates. At the configuration $(x_1,x_2)$, a single spin in each sub graph can either flip from up to down or from down to up. Therefore, there are four different terms in the master equation corresponding to leaving the current configuration, and similarly four transitions into the specific configuration:
\bal 
\nonumber
&\partial _t p(x_1,x_2) =
R^{u_1}(x_1-\Delta x,x_2)p(x_1-\Delta x,x_2) 
\\
\nonumber
&+ R^{u_2}(x_1,x_2-\Delta x)p(x_1,x_2-\Delta x) 
\\
\nonumber
&
+ R^{d_1}(x_1+\Delta x,x_2)p(x_1+\Delta x,x_2) 
\\
\nonumber
&+ R^{d_2}(x_1,x_2+\Delta x)p(x_1,x_2+\Delta x) 
\\
\nonumber
&
- \left[R^{u_1}(x_1,x_2)+R^{d_1}(x_1,x_2)\right.
\\
&\left.+R^{u_2}(x_1,x_2)+R^{d_2}(x_1,x_2)\right]p(x_1,x_2),
\eal
where $\Delta x$ is the change in the variable $x$ due to a single spin flip. In the limit $N \to \infty$  we approximate $x_1$ and  $x_2$ as continuous variables. Expanding both $p$ and all the terms of $R$ to first order in $\Delta x$ we get a Fokker-Planck like equation 
\bal
\label{eq:pFP}
\partial_t p &=  \partial_{x_1}[\left(R^{d_1}-R^{u_1}\right)p]+ \partial_{x_2}[(R^{d_2}-R^{u_2})p].
\eal
Note that in this case there is no diffusion, as the corresponding term vanishes in the $N\rightarrow\infty$ limit. Hence, it originates from a Langevin equation without random noise, namely from a deterministic equation for $x_1$ and $x_2$. For such deterministic motion, an initial distribution which is a delta function stays a delta function at all times, and it is therefore enough to know the evolution of the averages
\bal \label{eq:x12_average}
\overline{x_1}(t) \equiv \int x_1 p(x_1,x_2) dx_1 dx_2,
\\
\overline{x_2}(t) \equiv \int x_2 p(x_1,x_2) dx_1 dx_2.
\eal
Using these definitions, we write an ``equation of motion'' for the averages of $\overline{x_1}$ and $\overline{x_2}$ by substituting the values of the rates in \EQ{GlauberIsingRates} into Eqs.(\ref{eq:pFP}, \ref{eq:x12_average}). After some algebra these give: 
 \bal
 \dot{\overline{x_{1}}} &= \frac{1}{2}\left(  \tanh\frac{H-\overline{x_2}}{T_b}-\overline{x_1}\right), \nonumber\\
 \dot{\overline{x_{2}}} &=   \frac{1}{2}\left( \tanh\frac{H-\overline{x_1}}{T_b}-\overline{x_2}\right).
\label{eq:ave_x1_ave_x2}
 \eal
\OR{Unfortunately, these equations are not always linearly stable: for some values of $H,T_b,\overline{x}_1,\overline{x}_2$ a small perturbation in the initial values of $\overline{x_1},\overline{x_2}$  changes significantly the trajectory. For example, when the initial condition has  $\overline{x_1}=\overline{x_2}$ the symmetry of the dynamic keeps $\overline{x_1}$ and $\overline{x_2}$ equal at all times, even if the equilibrium distribution which corresponds to the specific $T_b$ and $H$ is different. In such cases,  any infinitesimal perturbation (in the initial condition or during the dynamic) results in relaxation towards the equilibrium, rather than following the solution of the above equations. Fortunately, in a large fraction of the parameter space $(H,T_b)$, as well as in the vicinity of all the fixed points, the above equations are stable. When stable, these non-linear equations describe the temporal evolution of the macroscopic system, and we can use them to understand the Mpemba behavior of the system.}\old{}
 
Using the above result, let us look at the equilibrium locus in the thermodynamic limit. For each value of $H$ and $T_b$, the equilibrium values of $\overline{x_1},\overline{x_2}$, which we denote by $\xi_1,\xi_2$,  are the steady state \EQ{ave_x1_ave_x2}, namely the solution of
  \bal
0&=  \tanh\frac{H-\xi_2}{T_b}-\xi_1, \nonumber\\
 0 &=   \tanh\frac{H-\xi_1}{T_b}-\xi_2.
\label{eq:x1_x2_equilib}
 \eal
 For each value of $H$, the equilibrium line can therefore be found using Eqs.(\ref{eq:x1_x2_equilib}) to calculate $\xi_1(T_b)$ and $\xi_2(T_b)$, for $0\leq T_b\leq \infty$. Note that  Eqs.(\ref{eq:x1_x2_equilib})  are symmetric to exchanging  $\overline{x_1}$ and $\overline{x_2}$. We therefore limit ourselves, without loss of generality, to \MV$\overline{x_1}\leq\overline{x_2}$.\old 
 
Examples for the equilibrium locus for $H=0.99$, $H=1$ and $H=1.01$ are shown in \FIG{fig_03_Equilibrium_Line_Different_H_v01.pdf}, where for each $T_b$ we have numerically found the equilibrium by solving \EQ{x1_x2_equilib}.  As expected, at $T_b\rightarrow \infty$ both $\xi_1\rightarrow 0$ and $\xi_2\rightarrow 0$. Importantly, the equilibrium line is not a simple convex line, therefore increasing the distance along the line does not necessarily increase the changes in the magnetizations. Moreover,  at $H=1$ the equilibrium locus has a singular transition demonstrated in \FIG{fig_03_Equilibrium_Line_Different_H_v01.pdf}. The sharp change in the equilibrium line at $H=1$ corresponds to the first order phase transition in the model at $H=1$ when $T_b=0$. This transition can be demonstrated by considering the limit $T_b\rightarrow 0$: for $H>1$,  the arguments of the hyperbolic tangents in \EQ{x1_x2_equilib} approach $+\infty$ asymptotically in the limit $T_b\rightarrow 0$, and hence  $\xi_{1,2} \rightarrow 1$.  In contrast, for $H<1$, $\xi_1\rightarrow1$ and $\xi_2\rightarrow -1$ are the asymptotic solutions at $T_b\rightarrow 0$.  As discussed in Sec.(\ref{sec:strong}), this sharp transition in the shape of the equilibrium line naturally corresponds to the sharp transition in the Mpemba-phases in \FIG{fig_02_phase_diagram_v01.pdf}: when the equilibrium locus abruptly changes, so does the coefficient along the slowest relaxation mode, $a_2(T)$.  Although this argument in principle should work only in the thermodynamic limit, it is evident from \FIG{fig_02_phase_diagram_v01.pdf} that in practice it works well already at $N=70$. 
 
 So far we have seen that some features of the finite system Mpemba phase diagram can be explained using the thermodynamic limit. In the next section we discuss the existence of  Mpemba effects in the thermodynamic limit, and their relations to the finite $N$ system.

\begin{figure}
\includegraphics[width=\linewidth]{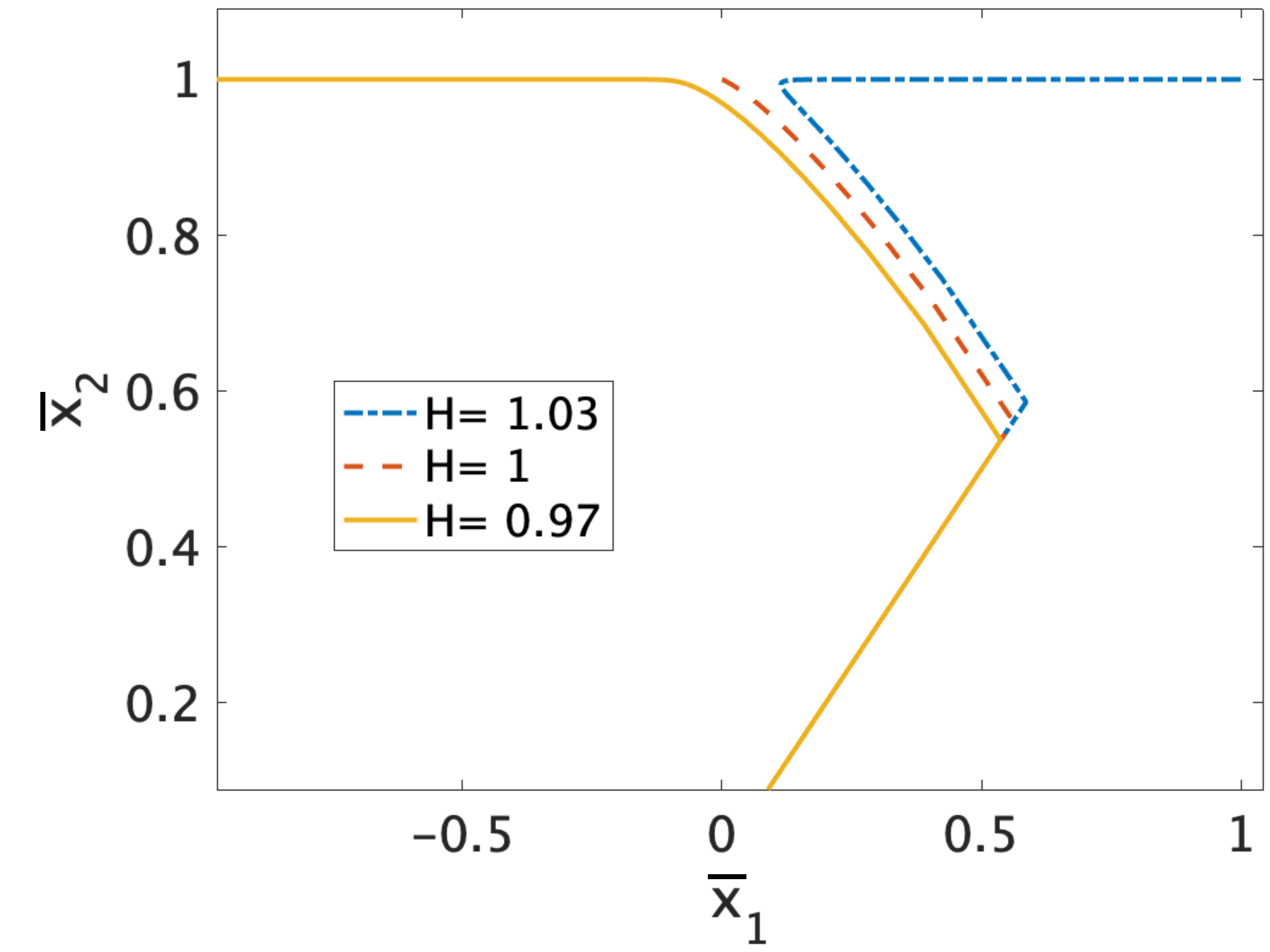}
\caption{\label{fig:fig_03_Equilibrium_Line_Different_H_v01.pdf}The ``equilibrium locus" of the mean-field anti-ferromagnet Ising mode at $H=0.97$, $H=1$ and $H=1.03$, plotted in the plain of average magnetization densities in each sub-graph $\overline{x_1}, \overline{x_2}$. Note the sharp transition in the curve's shape around $H=1$. This sharp transition corresponds to an abrupt transition in $a_2(T)$, which is clearly seen in the Mpemba phase diagram in~\FIG{fig_02_phase_diagram_v01.pdf}: the three equilibrium lines here corresponds to the three dashed lines in the lower panel of~\FIG{fig_02_phase_diagram_v01.pdf}.}
\end{figure}

 \subsection{Weak and Strong Mpemba effects in the Thermodynamic limit }
\label{Sec:Mpemba_in_Thermodynamic_limit} For systems with a finite number of states and a given set of environmental parameters, we have a simple prescription to check what types of  Mpemba effects exist: the monotonicity (weak effect) and zero crossings (strong effect) of the coefficient along the slowest dynamic, $a_{2}(T)$, encapsulate all this information. In the thermodynamic limit, we cannot use the same method, as rarely does the coefficient $a_{2}(T)$ have an analytic expression at finite $N$, for which we can take the thermodynamic limit, $N\rightarrow \infty$. Likewise, the direct calculation of $a_2(T)$ in the infinite system is often not feasible. This is somewhat unfortunate, because all the experimental observations of Mpemba effects mentioned above are in macroscopic systems, hence it is not clear that the mechanism we suggest is relevant for such systems. Moreover, the existence of the effect in macroscopic systems does not follow trivially from its existence in small systems. Although phase space becomes larger in large systems thus more shortcuts may exist, in the thermodynamic limit the probability distribution is concentrated in a tiny portion of the systems phase space, which suggests that the system would rarely explore the extended phase space, and therefore these shortcuts might not be as relevant. 

Although we cannot use $a_2(T)$ to analyze the existence of Mpemba effects in the thermodynamic limit of the antiferromagnet Ising model, for any  environmental conditions ($T_b,H$) and two temperatures $T_h$ and $T_c$, \EQ{ave_x1_ave_x2} can be used to compare the  relaxation trajectories initiated from the corresponding equilibrium distributions. A natural and physically motivated distance function in this case is the free energy difference between the current state and the equilibrium state, namely
\begin{eqnarray}\label{eq:DistFromEq}
D[(\overline{x_1},\overline{x_2}),(\xi_1,\xi_2)] &=& \mathcal{F}(\overline{x_1},\overline{x_2}) - \mathcal{F}(\xi_1,\xi_2),
\end{eqnarray}
where the free energy is given by \cite{Mean_field_MJP_1997unified}:
\begin{eqnarray}
\mathcal{F}(x_1,x_2) &=& \frac{\mathcal{H}(x_1,x_2) }{N} + \frac{T_b}{4}(1+x_1)\log(1+x_1)\nonumber\\
& &+ \frac{T_b}{4}((1-x_1)\log(1-x_1)\nonumber\\
& &+ \frac{T_b}{4}(1+x_2)\log(1+x_2)\nonumber\\
& &+ \frac{T_b}{4}((1-x_2)\log(1-x_2),
\end{eqnarray}
and $\mathcal{H}(x_1,x_2)$ is given by \EQ{Antiferro_hamiltonian}). \MV
If the initial condition with a longer distance from the equilibrium becomes, after some finite time, closer to equilibrium, then we know that there is a Mpemba effect in this system.  However, checking if a Mpemba effect exists at a point using this approach is tedious -- it requires solving the relaxation trajectories for all initial conditions.\old  Luckily, checking if strong effects exist and identifying their index is a much easier task. To this end we can linearize Eqs.(\ref{eq:ave_x1_ave_x2}) near the equilibrium point corresponding to $(T_b,H)$. Denoting the  differences from the equilibrium by $\Delta x_i = \xi_i - \overline{x_i}$, we can write for small $\Delta x_i$ 
\begin{eqnarray}
		 	\Delta\dot{x}_1 &=&  \frac{  \beta_b \left(1-\xi_1^2\right)\Delta x_2 - \Delta x_1}{2} + \mathcal{O}(\Delta x_i^2),\\
		 	\Delta\dot{x}_2 &=& \frac{  \beta_b \left(1-\xi_2^2\right)\Delta x_1 - \Delta x_2}{2}+ \mathcal{O}(\Delta x_i^2).
\end{eqnarray}
\MV These linearized equations have two relaxation eigen-directions -- a fast direction and a slow direction, with relaxation rates given by
\begin{eqnarray}\label{eq:linearized_lambda}
\bar \lambda_1 &=& -\frac{1}{2}\left(1-\beta_b\sqrt{(1- \xi_1^2)(1-\xi_2^2)}\right), \\
\bar \lambda_2 &=& -\frac{1}{2}\left(1+\beta_b\sqrt{(1-\xi_1^2)(1-\xi_2^2)}\right).
\end{eqnarray}
Unless the initial condition is such that at the long time limit the coefficient of points on its trajectory along the slow direction (namely the eigenvector corresponding to $\bar \lambda_1$) is zero, at long enough time the relaxation is from the direction corresponding to the slow direction. The number of trajectories that start on the equilibrium locus and approach the equilibrium point asymptotically from the fast direction is the Mpemba index, and the corresponding initial conditions  show strong Mpemba effect. To find if such initial conditions exist, we can shoot backwards in time a solutions to Eqs.(\ref{eq:ave_x1_ave_x2}) that approach the equilibrium from the fast direction (there are two such trajectories -- one from each side of the equilibrium locus). The number of crossing between these shoot-back trajectory and the equilibrium locus is the Mpemba index. 
 
As an example, consider the relaxation dynamic for environment with $H=1.1$ and $T_b=0.5$. The equilibrium locus as well as the relaxation trajectories from different initial temperatures are plotted in the upper panel of~\FIG{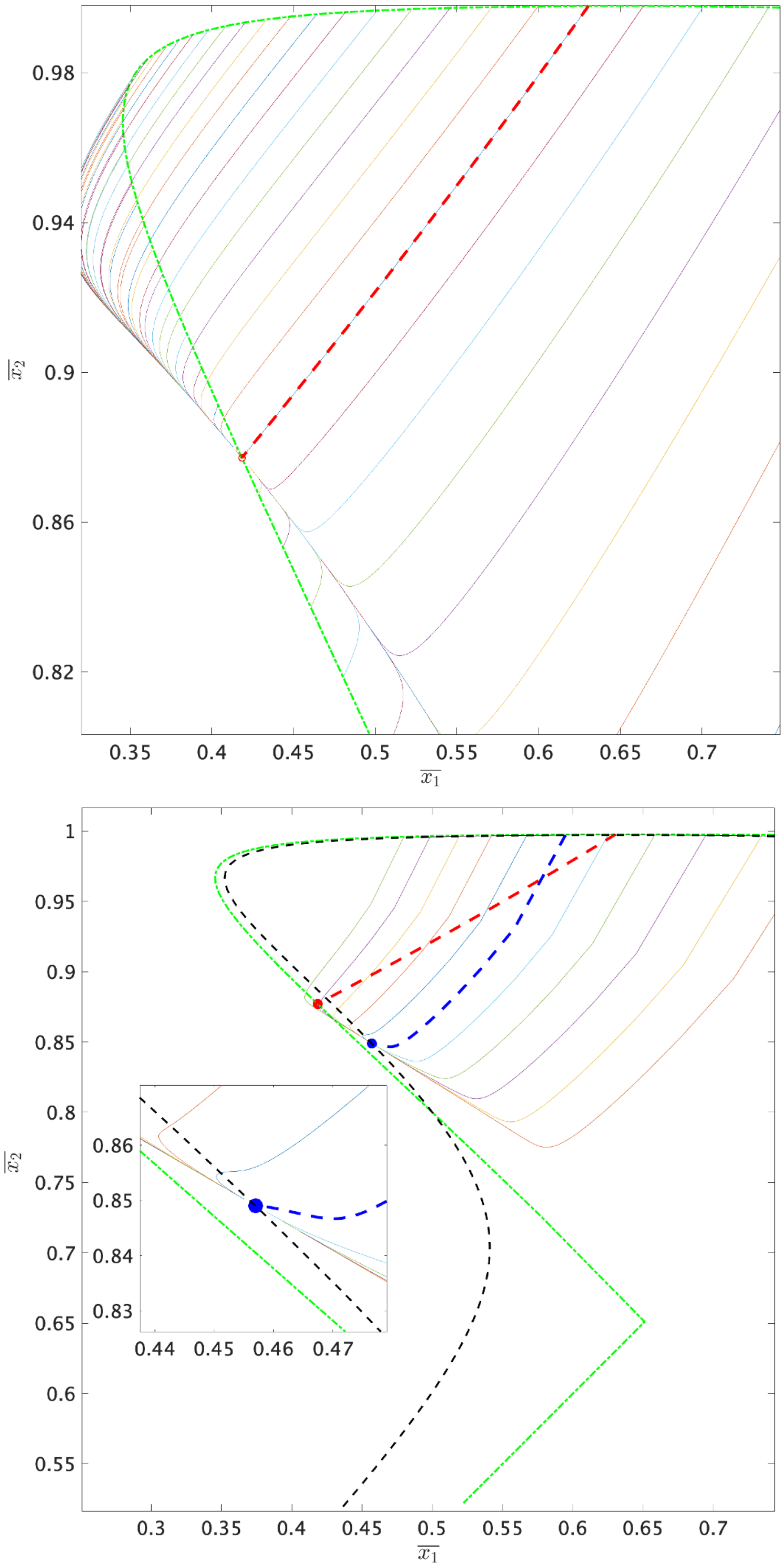}.
 \begin{figure}
\includegraphics[width=\linewidth]{fig_04_Relaxation_2_H1_v01.pdf}
\caption{\label{fig:fig_04_Relaxation_2_H1_v01.pdf} {\bf Upper panel:} The thick green dashed line is the \emph{equilibrium locus}, and the colored lines are the relaxation trajectories toward the equilibrium point (the red circle). All relaxation trajectories, except the dashed-red one, approach the equilibrium form the slow direction. The red dashed trajectory approach the equilibrium from the fast direction, and it corresponds to the strong inverse Mpemba trajectory. Its relaxation rate is 14.7 faster than the other trajectories -- see~\FIG{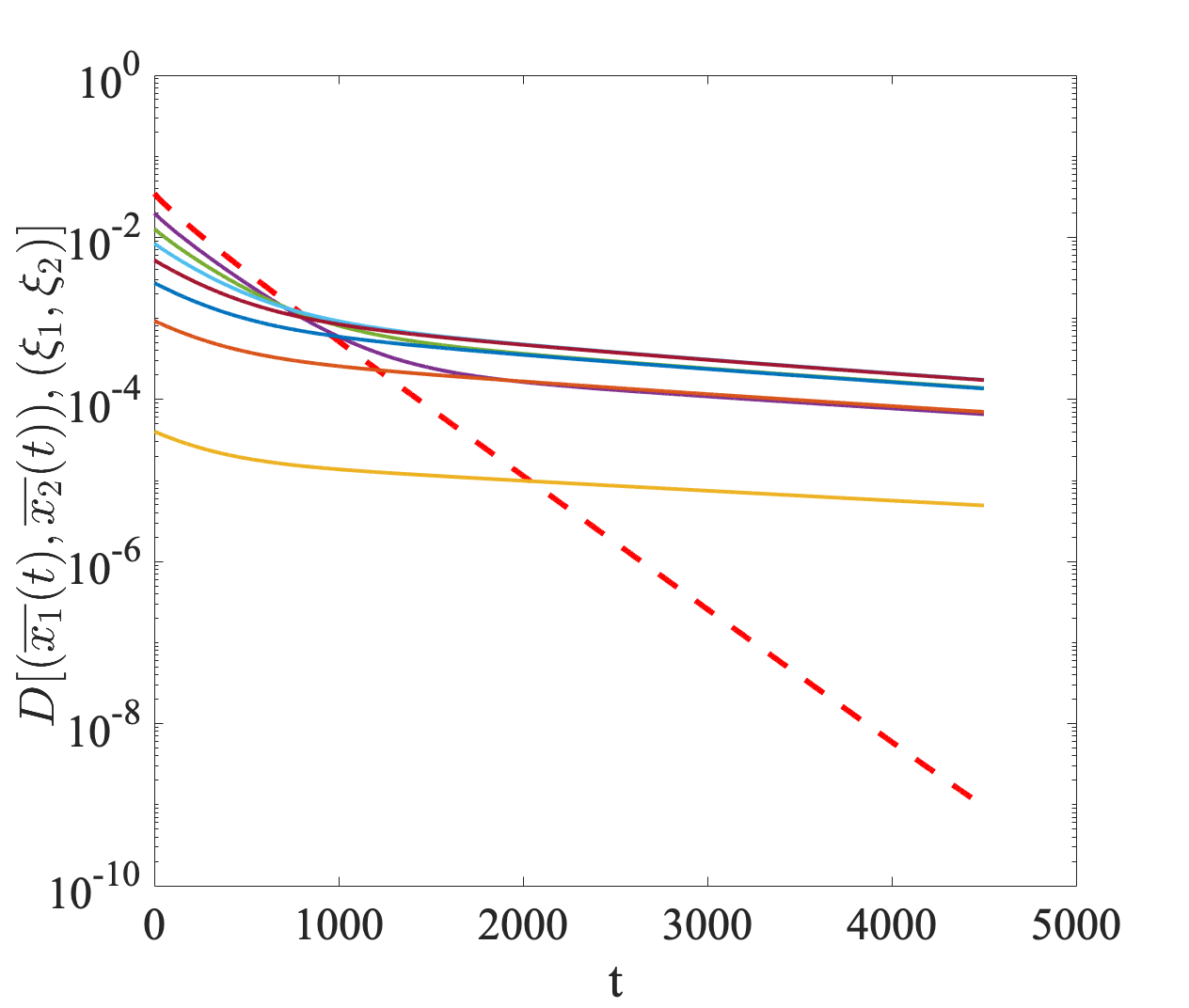}. {\bf Lower panel:} A comparison between the thermodynamic limit and finite system with $N=400$. The dash-dot thick green line is the thermodynamic limit equilibrium line, and the dashed black line is the projection into $(\overline{x_1},\overline{x_2})$ of the equilibrium line in the finite system. The projections of the relaxation trajectories for several initial temperatures are shown as thin lines, and the relaxation which corresponds to the strong Mpemba (calculated by $a(T_M)=0$) is in thick blue dashed line. The inset shows a blow-up of the relaxation trajectories near the equilibrium. The strong Mpemba initial condition approach equilibrium from a different direction.}
\end{figure}
 As can be seen in the figure, there is a ``fast'' and a``slow''  direction to the relaxation process, and essentially all the trajectories relax to equilibrium from the ``slow'' direction -- except for a single trajectory (the red dashed trajectory) that relax directly from the fast direction. This special trajectory is the strong  inverse Mpemba with $\mathcal{I}_M=1$ -- in agreement with the finite state phase diagram in \FIG{fig_02_phase_diagram_v01.pdf} for these values of $H$ and $T_b$. In this case, the ratio between the relaxation rates $\bar\lambda_1$ and $\bar\lambda_2$ given in \EQ{linearized_lambda} is 14.7. In other words, not only the strong  Mpemba initial condition relaxes exponentially faster, the relaxation rate is an order of magnitude higher that from any other initial temperature.  Indeed, as shown in \FIG{fig_05_DistVsTimeH1o1T0o5.png}, the strong effect trajectory relax significantly faster towards the equilibrium compared to all the other initial temperatures.
\old
 \begin{figure}
\includegraphics[width=\linewidth]{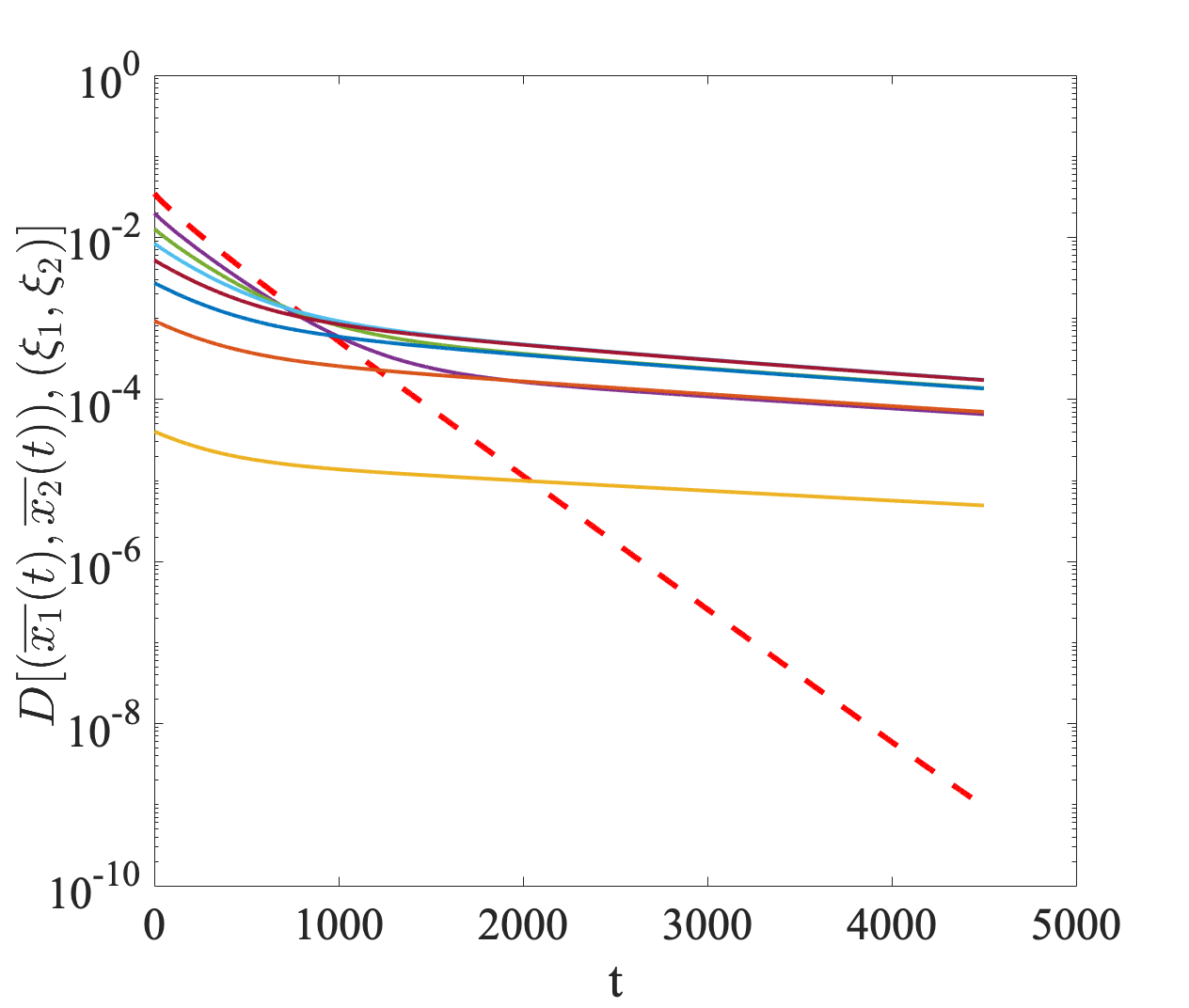}
\caption{\label{fig:fig_05_DistVsTimeH1o1T0o5.png}The distance from equilibrium (defined in~\EQ{DistFromEq}), in logarithmic scale, as a function of time for several initial conditions, in the thermodynamic limit. The initial condition corresponding to the strong effect is the dashed line.}
\end{figure}

\subsection{Comparing the thermodynamic limit with a finite $N$ system size}

\OR 
So far we have seen the strong Mpemba effect in both the finite state $N=400$ and the thermodynamic limit of the anti-ferromagnetic mean field Ising model. However, it is not obvious that the two effects are trivially related: In the thermodynamic limit the effect was derived by linearizing the non-linear equations for the order parameters, Eqs.(\ref{eq:ave_x1_ave_x2}), not by considering $a_2(T)$ in the $N\rightarrow\infty$ limit, which is intractable. One way to compare the two cases is to calculate a Mpemba phase diagram for the thermodynamic limit, and compare it with Fig.~\ref{fig:fig_02_phase_diagram_v01.pdf}. However, in the paramagnetic phase ($\overline{x_1}= \overline{x_2}$) we cannot use Eqs.(\ref{eq:ave_x1_ave_x2}) as they are not linearly stable, as discussed above. Therefore, we performed instead several other comparisons as presented next. 

One hint that the two effects are nevertheless related is given by the sharp transition of the equilibrium line at $H=1$ and the corresponding jump in the Mpemba index shown in \FIG{fig_02_phase_diagram_v01.pdf} and discussed above. To further convince that the strong Mpemba in the thermodynamic limit corresponds to the finite system case, the temperature at which a strong effect occur, $T_{M}(N)$, was calculated for various values of system size $N$. \FIG{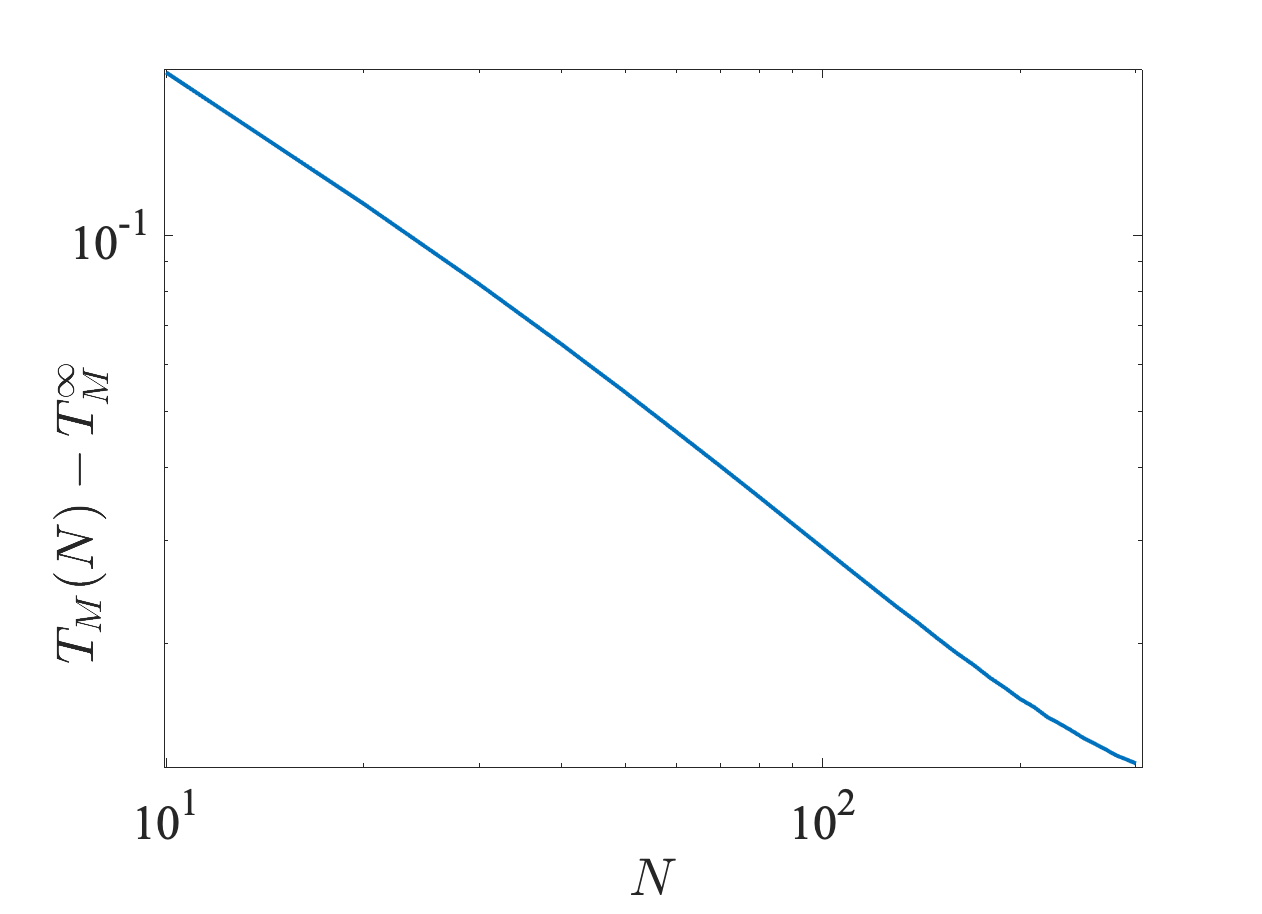} shows that indeed $T_{M}(N)$ converges, at the large $N$ limit, to the temperature $T^{\infty}_{M}$ at which a strong effect occur, in the thermodynamic limit, for the chosen $T_b$ and $H$. Similar behavior was observed in a wide range of temperatures and magnetic fields.

Additional comparison between a finite system and the thermodynamic limit is shown in the lower panel of \FIG{fig_04_Relaxation_2_H1_v01.pdf}. For $N=400$, we have calculated the equilibrium distribution as a function of the temperature, and ``projected'' it into the $(\overline{x_1},\overline{x_2})$ plane by calculating the equilibrium averaged magnetization in each sublattice. This equilibrium line is given by the thick black dashed line in the figure. For comparison, the green dash-dotted line is the thermodynamic limit equilibrium line, as in the upper panel. Next, we  calculated the probability distribution dynamics, initiated at several Boltzmann distributions of temperatures in the range $T\in[0.1,0.2]$. This was done using \EQ{MasterEq}, with $R$ defined in \EQ{GlauberIsingRates}. These trajectories in probability space were projected to the   $(\overline{x_1},\overline{x_2})$  plane, and are shown by the colored thin lines. Lastly, the trajectory of the strong effect $T_M=0.148$ that solves $a_2(T_M)=0$, was calculated too, and is plotted by the thick blue dashed line. For comparison, the strong Mpemba trajectory in the thermodynamic limit, initiated at $T_M^{\infty}=0.1377$, is plotted in thick dashed red line.  As shown in the inset, all the trajectories except the strong Mpemba approach the equilibrium point (the blue circle) from the ``slow'' direction, and the strong effect approaches the equilibrium point from a different -- ``fast'' direction. Note that the equilibrium line and the trajectories in the thermodynamic limit are not identical to those of the finite system, hinting that $N=400$ is not large enough to match the thermodynamic limit. Nevertheless, this example demonstrates that the $a_2(T_M)=0$ in finite size systems maps into the strong Mpemba mechanism in the thermodynamic limit. Although not a mathematical proof, this analysis hints that the strong Mpemba effect in the thermodynamic limit is a consequence of the strong effect in the finite system.
\old
 
\begin{figure}
\includegraphics[width=1\linewidth]{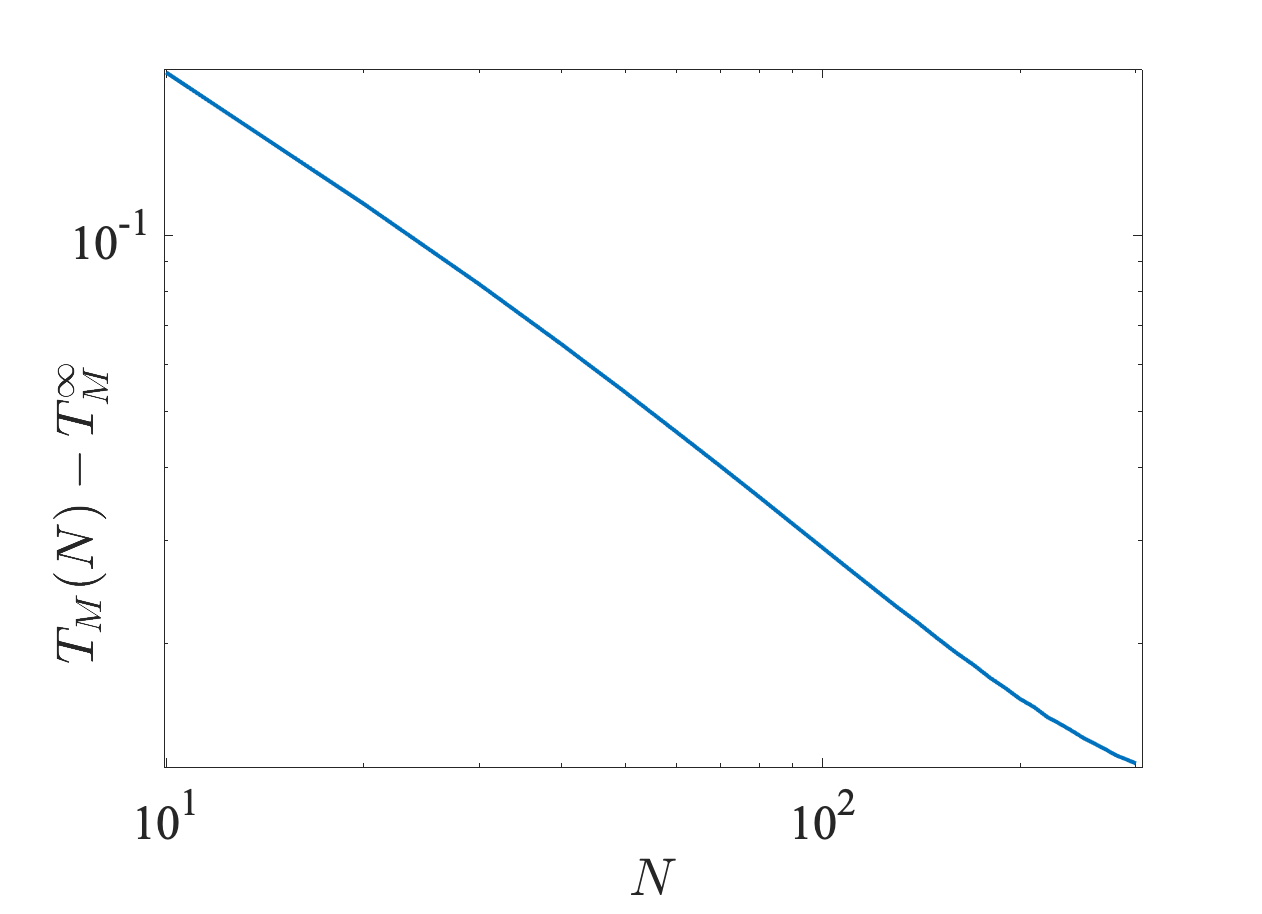}
\caption{\label{fig:fig_06_TminusTtmN300.png}Comparison between the strong effect in $N$ spin system and the strong effect in the thermodynamic limit. The temperature from which there is a strong effect at finite $N$ system, $T_{M}(N)$, converges to the temperature from which there is a strong effect in the thermodynamic limit, $T_{M}^{\infty}$. }
\end{figure}

\subsection{Temperature overshooting during Relaxation} 
\label{Sec:Overshoot}
As discussed above, the existence of a Mpemba effect can be checked by calculating the distance from equilibrium as a function of time from two different initial points on the equilibrium lines. This requires a choice of some reasonable distance function, e.g. the free energy distance (see~\EQ{DistFromEq}). In the specific case of the mean field anti-ferromagnet model there is another natural option: even though the system is not in equilibrium through the relaxation, it is possible to associate a temperature with each state during the relaxation process, and use this temperature to compare different relaxations. This temperature does not have all the properties commonly required from a distance function, e.g. it is not monotonically decreasing in a relaxation, nevertheless, it shed light on additional counter-intuitive aspect of thermal relaxations far from equilibrium: as shown below, the temperature can overshoot the environment temperature. In other words, a hot system, when coupled to a cold bath, can reach during its relaxation process  temperatures which are \emph{lower} than the environment's temperature. 

To associate a temperature for each point in the relaxation process, let us use the following coordinate transformation, from $(\overline{x_1},\overline{x_2})$ to  $(T_{eq},H_{eq})$, defined by:
 \bal\label{eq:Mapping_x12_HT}
 H_{eq} \equiv& 
 \frac{\overline{x_1}\tanh^{-1}\overline{x_1} - \overline{x_2}\tanh^{-1}\overline{x_2}}{\tanh^{-1}\overline{x_1}-\tanh^{-1}\overline{x_2} },\nonumber\\ 
T_{eq} \equiv& \frac{\overline{x_1}+\overline{x_2}}{\tanh^{-1}\overline{x_1}-\tanh^{-1}\overline{x_2}}.
 \eal
The physical significance of this transformation can be understood by a simple algebraic manipulation of the above equations that gives \EQ{x1_x2_equilib}. Comparing these to \EQ{ave_x1_ave_x2} one notes that for environment with temperature $T_b=T_{eq}$ and external magnetic field $H=H_{eq}$, the specific state given by $(\overline{x_1},\overline{x_2})$ is the equilibrium. In other words, if during the relaxation process, when the system is in the state $(\overline{x_1},\overline{x_2})$,  the system is decoupled from the current environment and coupled to a different environment with $T_b=T_{eq}$ and $H=H_{eq}$, then the system would be in equilibrium with the new environment. It is therefore natural to interpret $H_{eq}$ and $T_{eq}$ as the magnetic field and temperature of the system itself. 

Before proceeding, two comments on the above mapping are in place. (i) Note that the transformation is singular at  $\overline{x_1}=\overline{x_2}$ as the denominator  in Eqs.(\ref{eq:Mapping_x12_HT}) vanishes. In other words, we cannot associate a single temperature and magnetic field for states in which $\overline{x_1}=\overline{x_2}$. (ii) The ability to associate equilibrium temperature and magnetic field to most states of the system is a very non-generic property. It is a consequence of the fact that the number of parameters in the model is identical to the number of order parameters describing the system in the  thermodynamic limit. Luckily, in the thermodynamic limit of this model the probability distribution becomes a delta function with exactly two order parameters. 

Using the mapping in \EQ{Mapping_x12_HT}, we plotted in \FIG{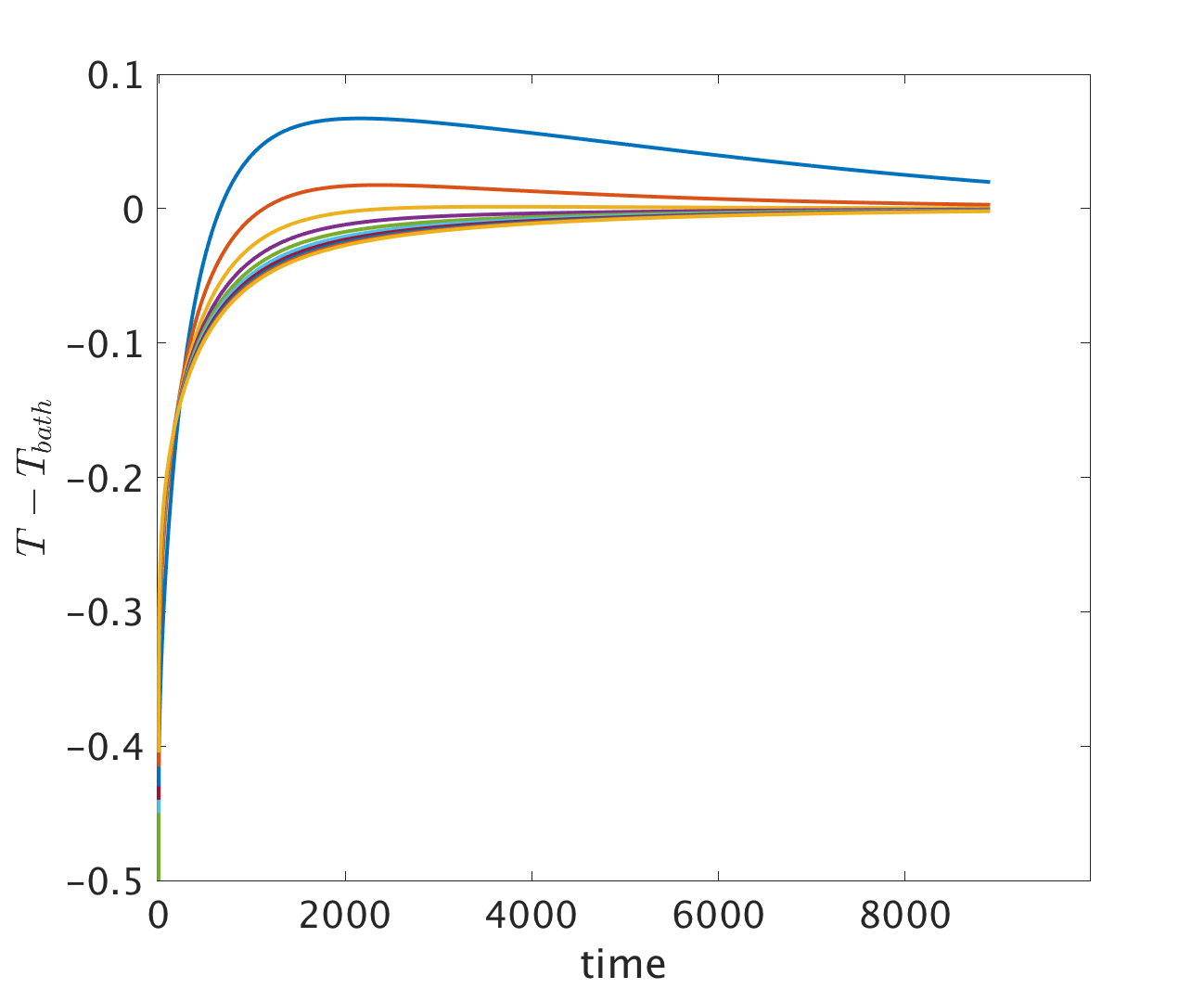} the temperature of the system as a function of time for various initial conditions along the equilibrium line. As can be seen, not only the temperature curves cross -- namely a Mpemba effect occurs,  but also for some relaxation trajectories the temperature is non-monotonic as a function of time. Moreover, systems that were initiated at temperatures lower than the environment's temperature  can reach temperatures which are higher than that of the environment in their relaxation. Similar non-monotonic relaxations were discussed in the context of non-Markovian thermal relaxation \cite{anomalous2015} or finite baths \cite{Parrondo2017heating}, but as far as we know this is the first example for such non-monotonic relaxation in Markovian dynamics and in the thermodynamic limit.

It is interesting to note that the temperature overshoot is tightly connected with the strong Mpemba effect. To explain this, let us examine~\FIG{fig_04_Relaxation_2_H1_v01.pdf} carefully. Initial temperatures to the left of the strong Mpemba relaxation trajectory approach equilibrium from   one side of the slow direction,  and initial temperatures to the right of the strong Mpemba  trajectory approach the equilibrium from the opposite direction, which is also a slow direction. Opposite directions mean opposite directions in the coordinate $T_{eq}$, namely there are trajectories that approach the equilibrium both from higher and lower temperatures compared to the environment. Conversely, if there are trajectories that approach equilibrium from both higher and lower temperature, they must approach equilibrium from opposite directions of the slow relaxation. Therefore, by continuity there must also be a trajectory that approach equilibrium from the fast direction - and this trajectory corresponds to a strong effect.
\begin{figure}
\includegraphics[width=1\linewidth]{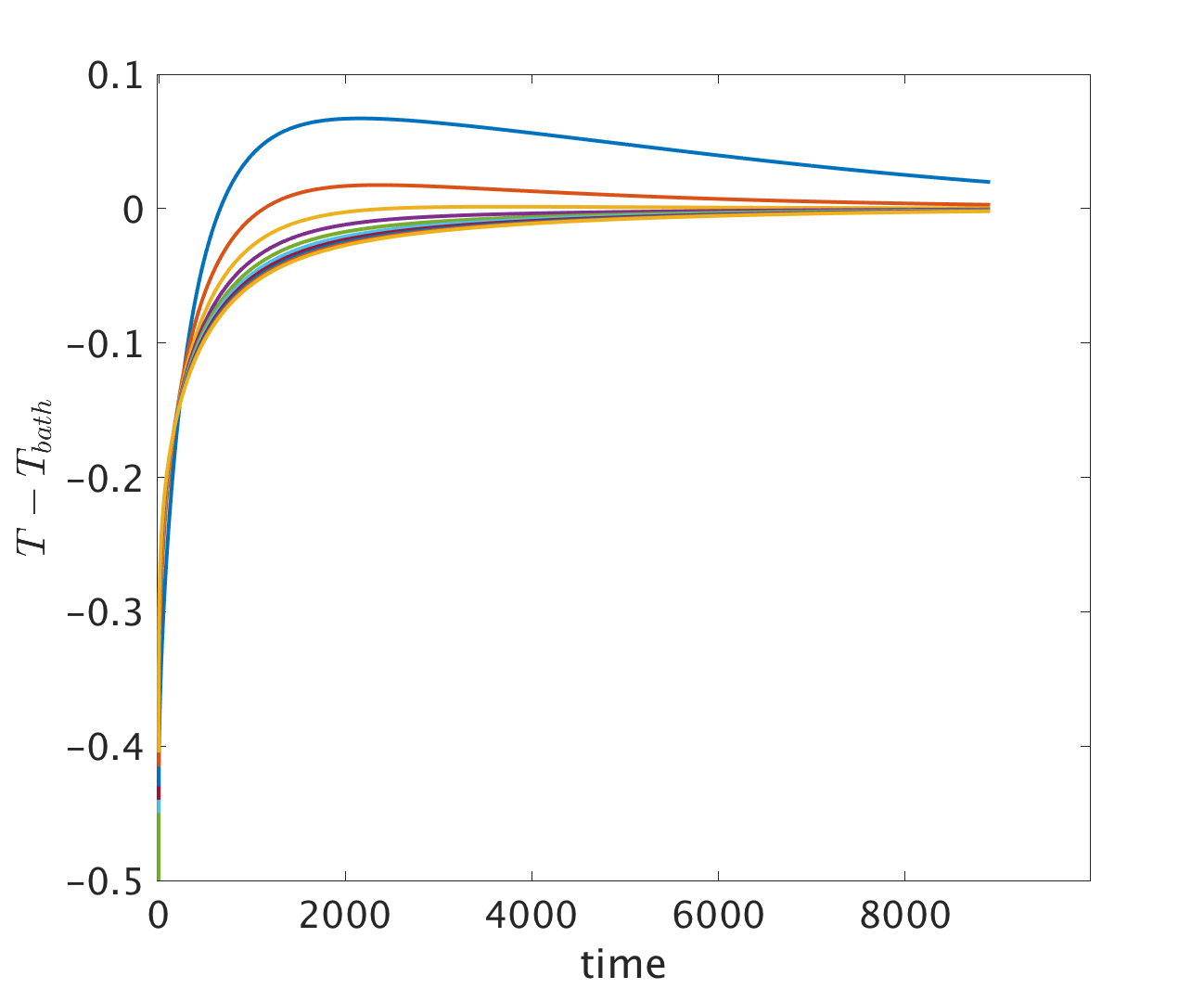}
\caption{\label{fig:fig_07_Temp_vs_time.png}$T_{eq}-T_{bath}$ as a function of time for various initial temperature along the equilibrium line. The inverse Mpemba effect is shown here as a crossing of two curves -- initially colder system heats up faster. The figure also shows that the temperature can overshoot the environment temperature -- the system can reach equilibrium temperatures which are higher than that of the environment.}
\end{figure} 

\section{How generic is the Mpemba effect in the REM model?} 
\label{subse:REM0}

So far, we considered the Mpemba effect in a specific model -- the mean field antiferromagnet Ising model. Is the existence of Mpemba effect a special property of this model, or should we expect to see similar effects in many other models? To address this question, we next evaluate the probability to have a Mpemba effect in a class of models with random parameters. As discussed below, the strong Mpemba effect plays a crucial role in our ability to estimate the probability of having a Mpemba effect. 

\subsection{Analytical estimates -- The Isotropic ensemble}
\label{subse:Isotropic_ensemble}

A proper analysis of the probability to find a Mpemba effect in classes of random models is a formidable challenge: one must perform the rather difficult calculation of the second eigenvector $\bm f_2$ and the coefficient $a_2$ in \EQ{a2} as a function of the initial temperature, the energies and the barriers and then average over the ensemble. Even the simpler problem of analyzing the strong Mpemba effect requires facing the daunting  task of analyzing the number of zeros in \EQ{a2}. 

To gain some analytical insight, we proceed by estimating the strong Mpemba probability in an ensemble of relaxation dynamics which we call \emph{the isotropic ensemble}. The ensemble is chosen to represent a wide distribution of barriers, so that the distribution of eigenvectors of the relaxation modes is as isotropic as possible consistent with a given target thermal distribution. \MV Explicitly, given a set of $L$ energies $\{E_{1},E_2,...,E_L\}$, we average over an ensemble of random $\bm f_{2}$ eigenvectors that are orthogonal, in the sense of the quadratic form in \EQ{a2}, to the equilibrium distribution with a given bath temperature $T_{b}$.\old This approach allows us to perform analytically the ensemble averaging. We then compare our analytic results for the isotropic ensemble with direct numerical calculations on the matrix \EQ{Driving} with fixed energies and random barriers, and find a surprisingly good agreement in certain parameter regimes. 

For a given set of energies $E_1, ..., E_L$ and dynamics prescribed by the Markov matrix \EQ{Driving} the steady state distribution is the Boltzmann distribution at $T_b$: $\bm \pi (T_b)$. The first eigenvector of the symmetrized Markov matrix $\tilde{R}$, is 
\bal
\bm f_{1}\equiv F^{1/2}\bm \pi(T_b) = \frac{1}{Z(T_b)}\left(e^{-\frac{\beta _b E_1}{2}},\hdots,e^{-\frac{\beta _bE_L}{2}}\right).\label{eq:fone}
\eal
and $\tilde{R}\bm f_{1} = \bm 0$. The second eigenvector of $\tilde{R}$, $\bm f_{2}$, together with the initial condition $\bm \pi(T)$, determine the coefficient $a_2$,  which according \EQ{a2} is
\bal
\label{eq:a2propto_0}
a_2(T)
=&\sum_{i=1}^L \frac{(\bm f_2)_i }{||\bm f_{2}||^2}\frac{e^{-\left(\beta-\frac{\beta _b}{2}\right)E_i}}{Z(T)}.
\eal
We obtain the explicit expression for the parity of the direct Mpemba index as a function of $\bm f_2$ by plugging  \EQ{a2propto_0} into \EQ{index_dir} and find
\bal
\nonumber
\mathcal{P}(\mathcal{I}^{dir}_M) = 
&\theta\left(\left[\sum   _{j=1}^L(\bm f_2)_je^{-\frac{\beta _bE_j}{2}} \left(\langle E\rangle _b-E_j\right)\right]\right.
\\ 
\label{eq:gugw_dir_0}
&
\times\left.\left[\sum _{i=1}^L (\bm f_2)_i e^{\frac{\beta_bE_i}{2}}\right]\right),
\eal
where $\langle E\rangle _b\equiv\sum_{i=1}^L E_ie^{-\beta_b E_i}/Z(T_b) $ is the average energy in equilibrium at $T_b$. We can represent
\EQ{gugw_dir_0} in the form 
\bal \label{eq:P_geometry}
&
\mathcal{P} (\mathcal{I}_M^{dir}) = \theta[(\bm f_2\cdot \bm u^{dir})(\bm f_2\cdot \bm w)],
\eal
with the vectors $\bm u ^{dir}$ and $\bm w$ defined as  
\bal
\label{eq:udiv2}
(\bm u ^{dir})_i& \equiv e^{\frac{\beta_bE_i}{2}},
\\
\label{eq:wvec}
(\bm w)_i& \equiv e^{-\frac{\beta _bE_i}{2}} \left(\langle E\rangle _b-E_i\right). 
\eal
Note that the vectors ${\bm u^{dir}}$ and ${\bm w}$ appearing in this form depend solely on the set of energies and on the bath temperature -- they are independent of the barriers. Moreover the form of \EQ{P_geometry} has a simple geometric meaning. To see it we single out the components of $\bm f_2$ in the plane spanned by the (non-orthogonal) vectors ${\bm u^{dir}},{\bm w}$. 
Choosing $f_{\parallel}$ as the component of $\bm f_2$ parallel to $\bm u^{dir}$, we have:
\bal
\nonumber
\bm f_2=&f_{\parallel}  \frac{\bm u ^{dir}}{\|\bm u ^{dir}\|}+f_{\perp}\frac{ \left(\bm w-\frac{\bm w\cdot
   \bm u ^{dir}}{ \|\bm u ^{dir}\|^2}\bm u ^{dir}\right)}{\sqrt{ \|\bm w\|^2-\frac{(\bm w\cdot
   \bm u ^{dir}_2)^2}{ \|\bm u ^{dir}\|^2}}}
\\
&+\text{ terms orthogonal to } \bm u ^{dir} \text{ and }\bm w.
\eal
In this basis $(\bm f_2\cdot \bm u^{dir})(\bm f_2\cdot \bm w)$ is equal to
\bal 
\label{eq:conditiononf2}
& (\bm f_2\cdot \bm u^{dir})(\bm f_2\cdot \bm w)=\\ & f_{\parallel}^2 (\bm u^{dir}\cdot \bm w)+f_\parallel f_\perp |\bm u^{dir}\cdot \bm w|K(\bm u^{dir}, \bm w), \nonumber
\eal
where 
\bal
\label{eq:Kexpression}
K(\bm u^{dir}, \bm w)\equiv
   \sqrt{\frac{ \|\bm u^{dir}\|^2 \|\bm w\|^2}{(\bm w\cdot \bm u^{dir})^2}-1}.
\eal
Therefore on the $f_\parallel,f_\perp$ plane, the region satisfying $\mathcal{P} (\mathcal{I}_M^{dir}) \neq 0$ is a double wedge
\bal
\label{eq:lower_bound_explicit_0}
f_{\parallel}^2 (\bm u^{dir}\cdot \bm w)+f_\parallel f_\perp |\bm u^{dir}\cdot \bm w|K > 0,
\eal
(see \FIG{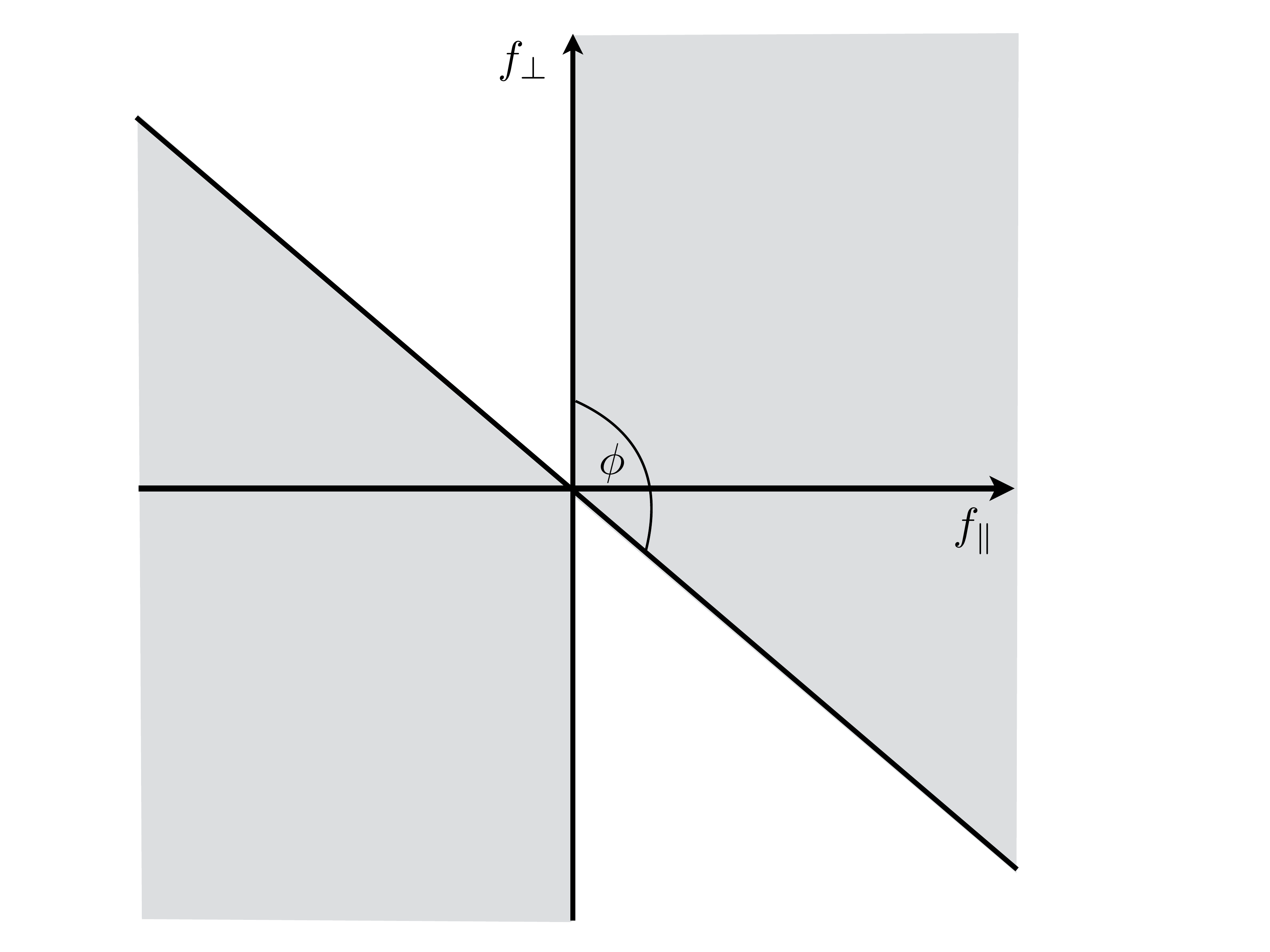}). The boundary of the region is associated with the lines $f_\perp=-f_\parallel/K$ and  $f_\parallel=0$.

The same treatment is also possible for the inverse Mpemba effect. For example, assuming, for simplicity, a non-degenerate ground state, and ordering the energies so that $E_1$ is the ground state energy, we find that $\mathcal{P}(\mathcal{I}^{inv}_M)$ is given by \EQ{P_geometry} with the replacement
\bal 
{\bm u^{dir}}\longrightarrow {\bm u^{inv}}~~~;~~~({\bm u^{inv}})_i= -e^{\frac{\beta_bE_i}{2}}\delta_{i,1}.
\eal

Next, we formulate the averaging over the admissible $\bm f_{2}$ vectors. In the class of random relaxations we consider, we generate $\bm f_{2}$ by picking a random vector $\bm g=(g_{1},...,g_L)$, and obtaining from it a random vector orthogonal to  $\bm f_{1}$ (by subtracting the projection of $\bm g$ on $\bm f_1$)
\bal\label{eq:isog}
\bm f_{2}(\bm g)\equiv \bm g-\frac{\bm g \cdot \bm f_{1}}{||\bm f_1||^2}\bm f_{1}.
\eal
The distribution of the $\bm g$ vectors is taken to be isotropic and therefore the projection of the distribution of $\bm g$s onto the hyperplane orthogonal to $\bm f_{1}$ is also isotropic. For this purpose, we take the $g_{i}$s in $\bm g$ to be IID Gaussian variables. Analogously to the derivation of, \EQ{P_geometry}, we plug \EQ{isog} into \EQ{gugw_dir_0} and separate the $g_i$ components. We find that the direct Mpemba parity, for a particular realization of $g_i$, can be written as
\bal
\mathcal{P} (\mathcal{I}_M^{dir}) = \theta[(\bm g\cdot \bm {u}_{iso}^{dir})(\bm g\cdot \bm w)],
\eal
with $\bm w$ is defined in \EQ{wvec} and $\bm {u}_{iso}^{dir}$ given by 
\label{eq:udir}
\bal  &(\bm {u}_{iso}^{dir})_i= e^{\frac{\beta_bE_i}{2}}-\frac{Le^{-\frac{\beta_bE_i}{2}}}{Z(T_b)},
\eal
where $L$ is the number of energy levels (or the system size). As before, we break $\bm g$ into the component parallel and perpendicular to $\bm {u}_{iso}^{dir}$ and find, as before:
\bal 
\label{eq:lower_bound_explicit}
&(\bm g\cdot \bm {u}_{iso}^{dir})(\bm g\cdot \bm w) 
\\
&= g_{\parallel}^2 (\bm {u}_{iso}^{dir}\cdot \bm w)+g_\parallel g_\perp |\bm {u}_{iso}^{dir}\cdot \bm w|K(\bm {u}_{iso}^{dir},\bm w),
\eal
where $K$ is defined in \EQ{Kexpression}. 
 
\begin{figure}[h]
\includegraphics[width=0.5\textwidth]{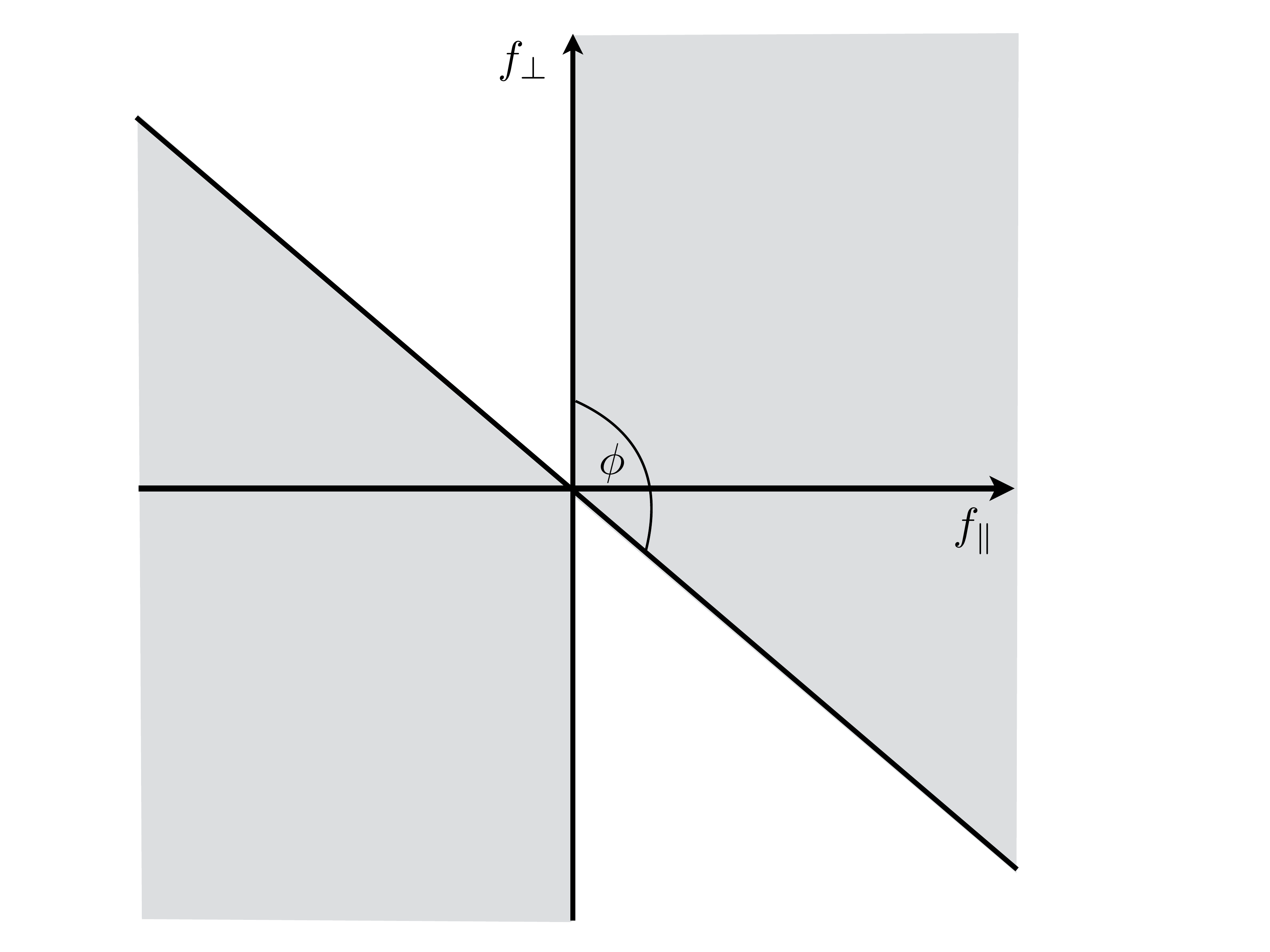}
\caption{\label{fig:fig_08_wedge_v03.pdf}The direct Mpemba index $\mathcal{I}^{dir}_M$ is odd in the double wedge shaded region of $f_\parallel f_\perp$ plane if $\bm u^{dir}_2\cdot \bm w>0$ and while if $\bm u^{dir}_2\cdot \bm w<0$ the direct Mpemba index is odd for the complementary region (white). See \EQ{lower_bound_explicit_0}.}
\end{figure}
 
The Gaussian IIDs $g_i$ have a rotationally-invariant joint distribution function, and therefore in any coordinate system the components are corresponding Gaussian IIDs. It follows that $g_\parallel, g_\perp$ are Gaussian IIDs and have a rotationally-invariant distribution on the $g_\parallel, g_\perp$ plane. On this plane, the region satisfying $\mathcal{P}(\mathcal{I}_M^{dir}) >0$ is a double wedge (c.f. \FIG{fig_08_wedge_v03.pdf}), and the probability of $g_\parallel, g_\perp$ to fall inside the wedge only depends on the wedge angle.

Geometrically, if $\phi$ is the angle between $\bm u$ and $\bm w$, then $\text{Prob}(\mathcal{P} (\mathcal{I}^{dir}_M)>0)=\frac{\phi}{\pi}$ 
when  $(\bm u\cdot \bm w)>0$ (and $\text{Prob}(\mathcal{P} (\mathcal{I}^{dir}_M)>0)=1-\frac{\phi}{\pi}$ when $(\bm u\cdot \bm w)<0$). Expressed explicitly in terms of $\bm {u}_{iso}^{dir}, \bm w$ we find:
\bal
\nonumber
&\text{Prob}(\mathcal{P}(\mathcal{I}_M^{dir})  > 0)=\frac{1}{2}
\\
\label{eq:conditionprob}
&+\frac{{ \text{sign}(\bm {u}_{iso}^{dir}\cdot \bm w)}}{\pi}\arctan \frac{1}{K(\bm {u}_{iso}^{dir},\bm w)}.
\eal
To recap, the formula \EQ{conditionprob} represents, for a given set of energies $\{E_i\}$ and bath temperature $T_b$, the probability that the direct Mpemba index is odd. 

\EQ{conditionprob} can be simplified for hot bath temperatures, $k_BT_b\gg {\rm max}(\{E_1,...,E_L\})$, and asymptotically it gives
\bal
\label{eq:HighTemp}
\text{Prob}(\mathcal{P}(\mathcal{I}_M^{dir})  > 0)\approx \frac{C_E}{T_{b}}.
\eal
Here the constant $C_{E}$ depends only on the first few moments of the energy level distribution (for the explicit expression see the appendix ~\ref{sec:high_temp}).

\FIG{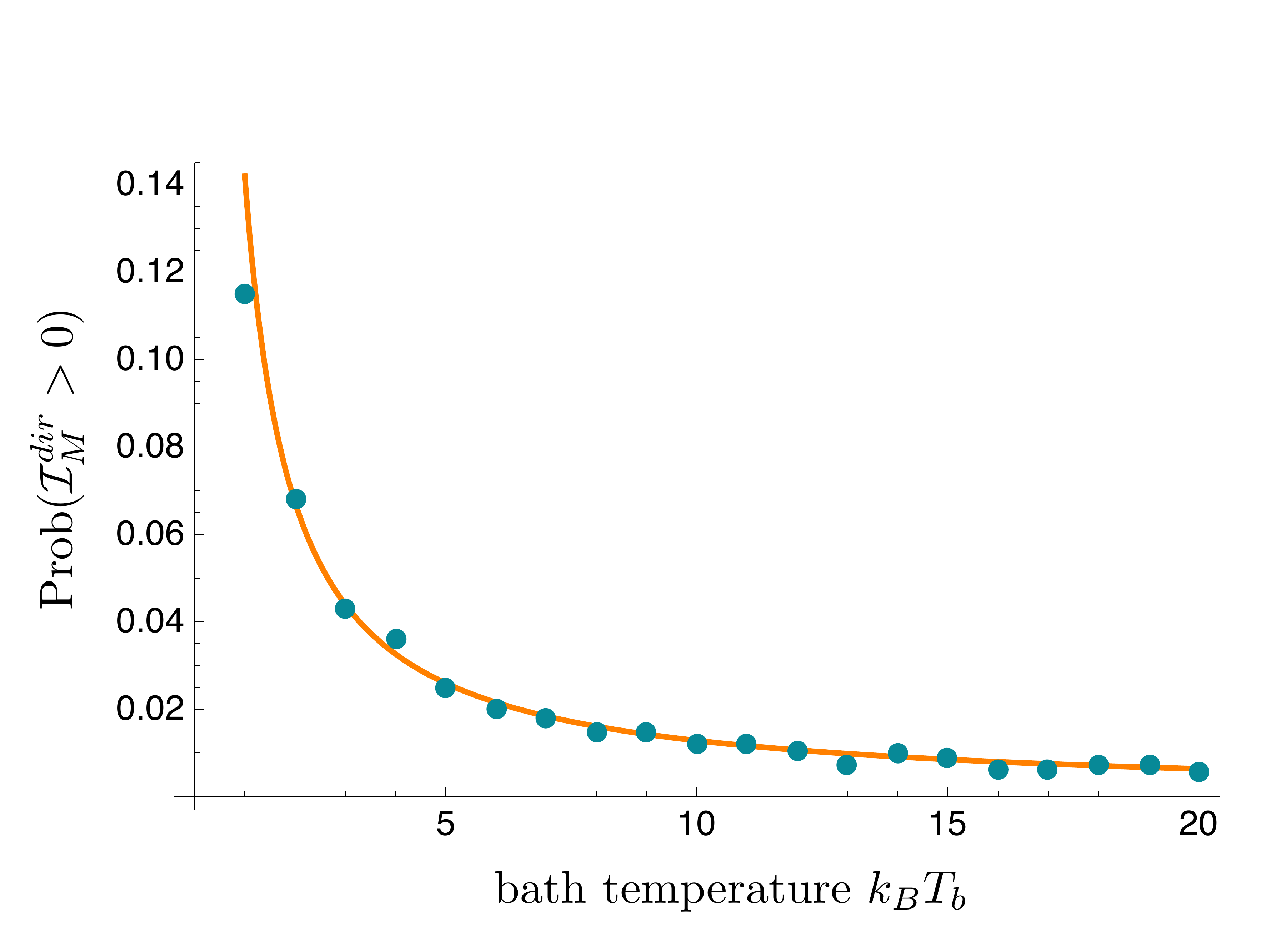} shows a comparison of \EQ{conditionprob} with a random realization of $L = 15$ energies, and Mpemba index averaged over $~4000$ realizations of random barriers. The expression seems to capture surprisingly nicely the behavior of for a random draw of energy levels when the barriers distribution is wider than the distribution of energies, and the temperature is higher than the characteristic energy spread.

Similarly, for inverse Mpemba we have
\bal
\mathcal{P} (\mathcal{I}_M^{inv}) = \theta[(\bm g\cdot \bm {u}_{iso}^{inv})(\bm g\cdot \bm w)],
\eal
with $\bm w$ is defined in \EQ{wvec} and $\bm {u}_{iso}^{inv}$ given by 
\bal 
\label{eq:uinv}
&(\bm {u}_{iso}^{inv})_i= -e^{\frac{\beta_bE_i}{2}}\delta_{i,1}+\frac{e^{-\frac{\beta_bE_i}{2}}}{Z(T_b)}, 
\eal
where we assumed that $E_1$ is the lowest energy. As before,
\bal
\nonumber
&\text{Prob}(\mathcal{P}(\mathcal{I}_M^{inv})  > 0)=\frac{1}{2}
\\
\label{eq:conditionprobInv}
&+\frac{{ \text{sign}(\bm {u}_{iso}^{inv}\cdot \bm w)}}{\pi}\arctan \frac{1}{K(\bm {u}_{iso}^{inv},\bm w)}.
\eal
Substituting for $\bm w$ and $\bm {u}_{iso}^{inv}$ from \EQS{wvec}{uinv} we get 
\bal
\mathcal{P}(\mathcal{I}_M^{inv}) = \frac{1}{2} - 
\frac{1}{\pi}
\arctan\left(\frac{1}{\sqrt{\frac{(Z(T_b)e^{\beta_b E_1}-1)\Delta E_b^2}{(E_1 - \langle E \rangle _b)^2}-1}}\right). 
\eal

The above expression simplifies in the limit of a very low bath temperature, $k_BT_b\ll (E_2 - E_1)$. Without loss of generality we set $E_1 = 0$ and obtain
\bal
\nonumber
&\text{Prob}( \mathcal{P}(\mathcal{I}_M^{inv})>0)\approx 
\frac{1}{2} 
\\
&- \frac{1}{\pi}\arctan\left(\sqrt{\frac{(E_2\varepsilon_{2} + E_3 \varepsilon_{3})^2}{(E_2 -E_3)^2\varepsilon_{2}\varepsilon_3(1 + \varepsilon_{2} + \varepsilon_3)}}\right). 
\eal
Simplifying the expression even further we get
\bal
\label{eq:Pinv_lowTb_3levels}
\text{Prob}( \mathcal{P}(\mathcal{I}_M^{inv})>0)\approx 
 \frac{1}{\pi} \sqrt{\frac{(E_2 -E_3)^2\varepsilon_{2}\varepsilon_3(1 + \varepsilon_{2} + \varepsilon_3)}{(E_2\varepsilon_{2} + E_3 \varepsilon_{3})^2}}, 
\eal
where $\varepsilon_{i} \equiv e^{-\beta_b E_i}$. Taking $\varepsilon_3 \to 0$ or if $E_2 = E_3$ we find 
\bal
\text{Prob}(\mathcal{P}(\mathcal{I}_M^{inv})>0)&\approx 
0,
\eal
which is expected as there is no Mpemba effect for a two level system.

It is important to note that the isotropic ensemble, while introduced for the purpose of enabling analytical averaging, is consistent with the assumptions of our relaxation dynamics. Namely, one can prove the following theorem.

{\it Theorem:} Given any choice of a real vector $\bm f_{2}$ orthogonal to $\bm f_{1}$ in \eqref{eq:fone}, there exists a set of barriers $B_{ij}$ with relaxation dynamics obeying detailed balance \eqref{eq:Driving} having $F^{-1/2}\bm f_{2}$ as its slowest relaxation eigenvector. 

The proof can be found in the appendix~\ref{sec:proof}.

\begin{figure}[h]
\includegraphics[width=\linewidth]{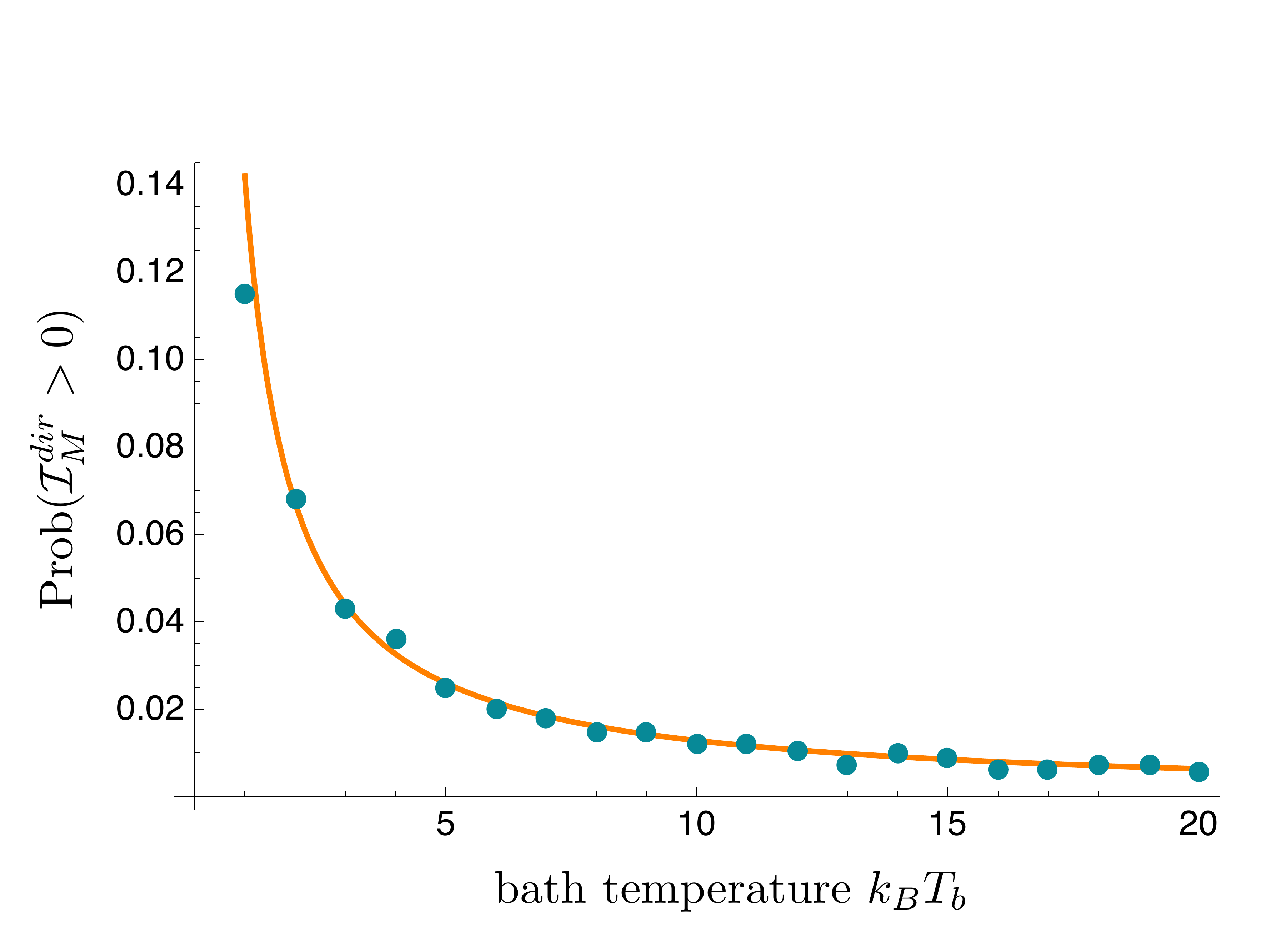}
\caption{\label{fig:fig_09_10Estd_edited_v01.pdf} The probability for odd direct Mpemba index for a particular random draw of $L=15$ energy levels obtained on two independent ways -- analytically by averaging over the \emph{isotropic ensemble} (solid line) and numerically by averaging over random barriers (points). The energy level realization was drawn from a Gaussian distribution with zero mean and standard deviation $1.5$. Each point corresponds to the average of $4000$ barrier realizations, taken from a  truncated Gaussian distribution with zero mean and standard deviation $15$ (The Gaussian was truncated to have only positive $B_{ij}$ values). The solid line represents the analytical result for the {\it isotropic ensemble} \EQ{conditionprob}.}
\end{figure} 

\subsection{\label{subse:REM0} Numerically -- the probability of the strong Mpemba effect in the REM model}

In the previous section we analyzed an analytically tractable model, however, in general, estimating the probability of the Mpemba effect is a daunting and often impossible task. In what follows, we numerically study the probability of having a strong Mpemba effect in the random energy model (REM). 

The random energy model was introduced by Derrida as an extreme limit of spin glasses~\cite{DERRIDA198029}. It is the simplest model of a system with quenched disorder that has a phase transition. In the REM, $L$ energy levels are IID random variables. The conventional choice for the probability distribution of $E_j$'s is a Gaussian distribution
\bal
\label{eq:pd_energies}
{\rm Prob}(E_j = E) = \frac{1}{\sqrt{2\pi\sigma^2_{E}\log_2L}}e^{-\frac{E^2}{2 \sigma^2_{E}\log_2L}}, 
\eal
where in order to have extensive thermodynamic potentials the variance depends on the system size. At temperatures lower than $T_{critical} \equiv \sigma _E /(k_B\sqrt{2 \ln 2})$ the system is trapped in a few low-lying states; this condensation phenomena is a phase transition and at the transition the free energy is non-analytic. 

\MV 
Note that quenched disorder is not necessary for the Mpemba effect. Our previous examples, such as the mean field Ising antiferromagnet and other examples discussed in~\cite{2016LuRaz} are a proof that quenched disorder is not an essential feature for this effect. 
\old

The Mpemba effect is a property of the system and its dynamics, thus to study it we need to specify the barriers $B_{ij}$ in (\ref{eq:Driving}). Here we chose $B_{ij}$ as IID random variables obeying a "truncated" Gaussian distribution
\bal
\label{eq:pd_barriers}
{\rm Prob}(B_{ij}  = B) = 
\frac{1}{\sqrt{2\pi\sigma^2_{B} \log_2L}}e^{-\frac{B^2}{2\sigma^2_{B}\log_2L }}\theta (B),
\eal 
and $\theta$ is the Heaviside step function. This particular choice of barriers can only impede the transition rates $R_{ij}$, as in this case $e^{-\beta _b B_{ij}} < 1$. The variance of the barriers is scaled with the system size like that of the energies, so that their ratio is system size independent. Note that numerous other choices of the dynamics for the REM have been studied in the past, most notably single spin flip dynamics (see e.g.~\cite{AgingREM2002,BenArousLesHouches,TrapRemDynamics2017} and references therein) and it would be interesting explore for Mpemba effects those other choices of REM dynamics as well.

Numerically we studied the parity of the direct Mpemba effect (see \EQ{index_dir}), by exact diagonalization of an ensemble of REM with random barriers $R$ matrices. 
As an example of typical numerical results, see~\FIG{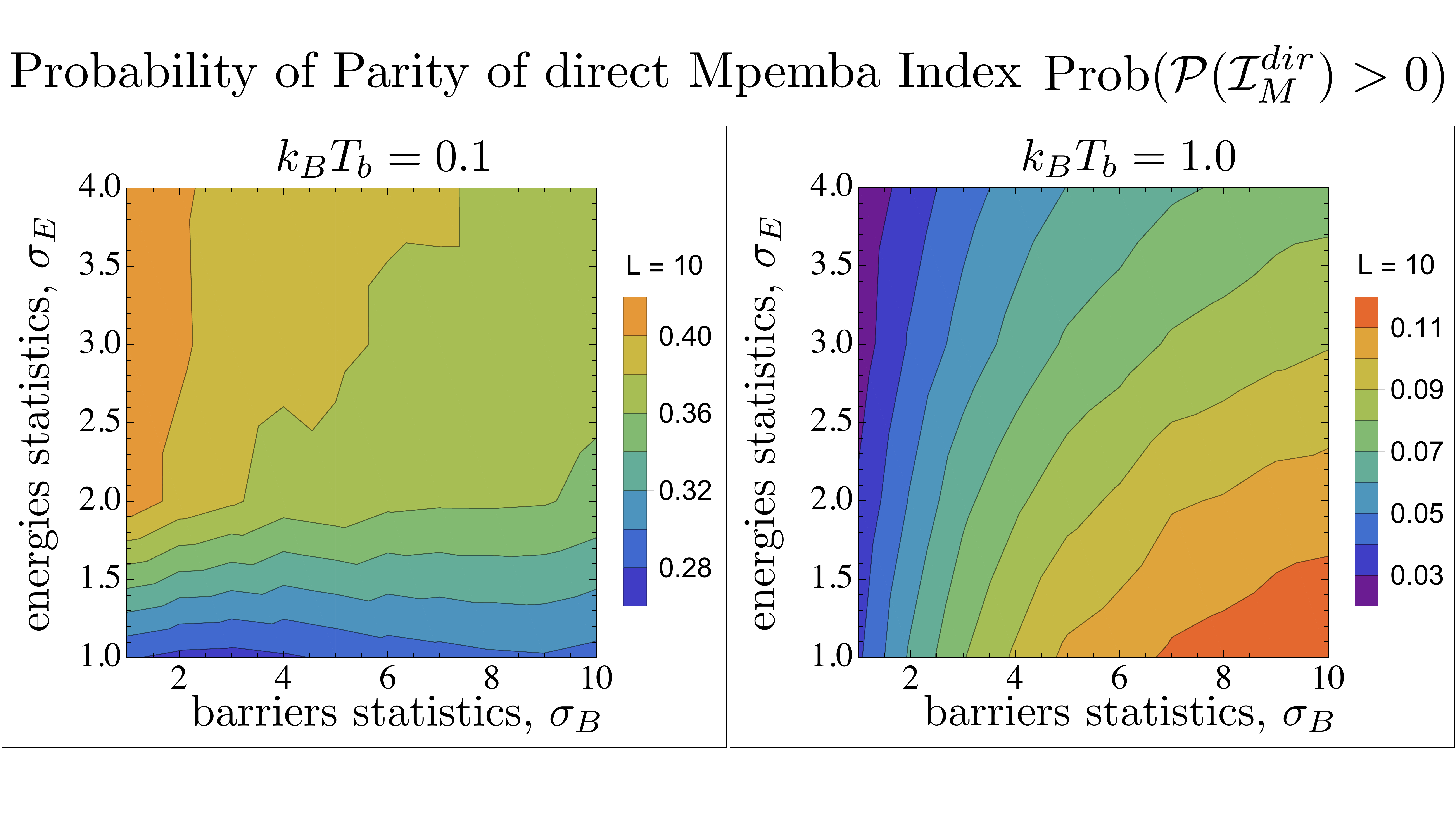}, where $L=10$ energy levels were chosen from a Gaussian distribution \EQ{pd_energies} and barriers chosen from \EQ{pd_barriers}. The bath temperature in the numerics was $k_BT_b = 0.1$ and $k_BT_b = 1.0$. Each data point was averaged over $10^5$ realizations. From the ample numerical evidence we infer that the Mpemba effect occurs with finite probability, especially for $T_b < T_{critical}$ case (left panel~\FIG{fig_10_REM_randbarrier_L10_M10000_Tbdiff_v01.pdf}).   
\begin{figure}[h]
\includegraphics[width=\linewidth]{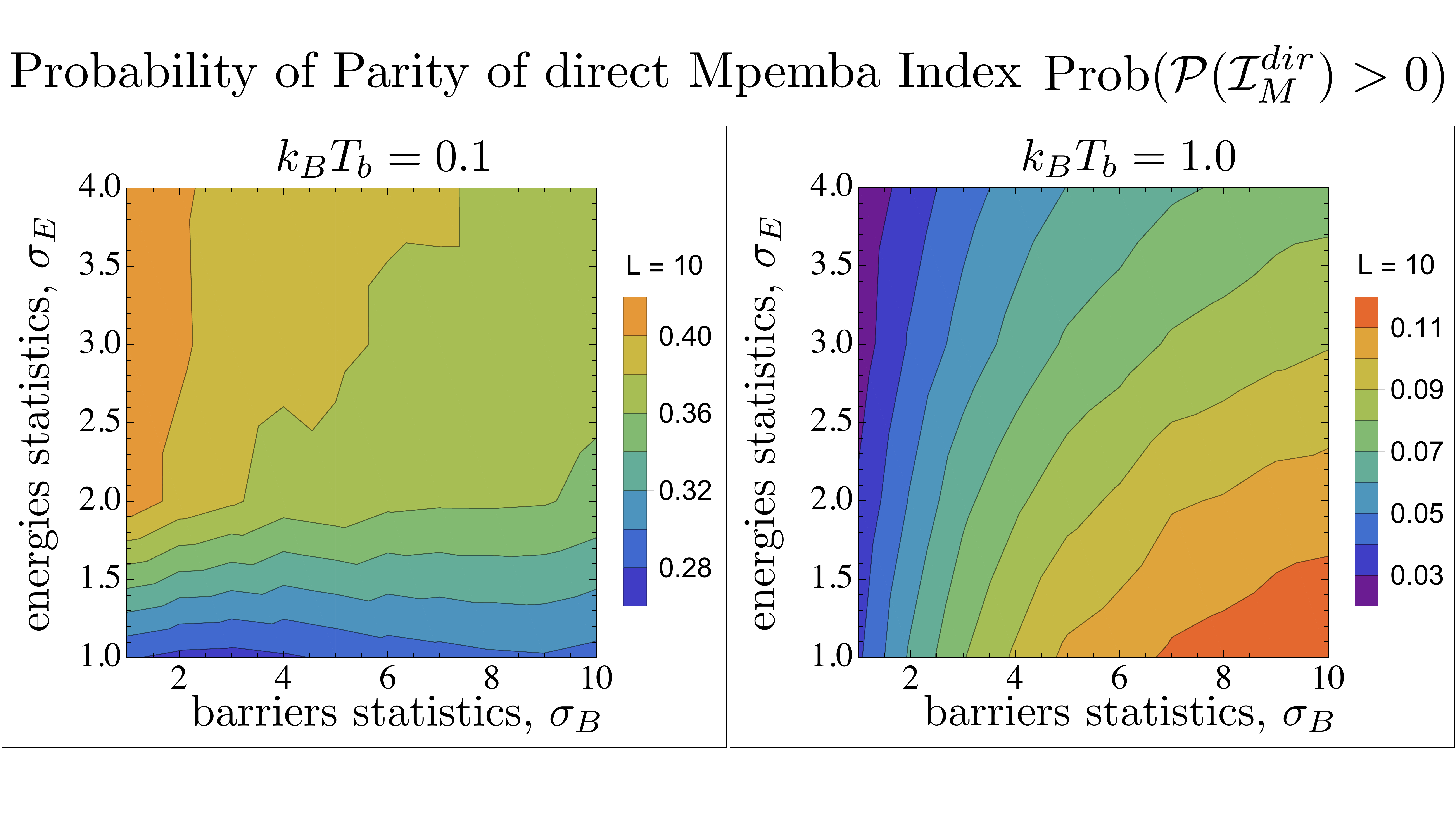}
\caption{\label{fig:fig_10_REM_randbarrier_L10_M10000_Tbdiff_v01.pdf} Lower bound for the probability of the strong Mpemba effect -- the probability of the parity being positive for the direct Mpemba index (see \EQ{index_dir}) for the case of REM with random barriers. The number of energy levels is $L = 10$ and bath temperatures are $k_BT_b =0.1$ (left) and $k_BT_b =1.0$ (right). The energies were IIDs from \EQ{pd_energies} with variance $\sigma_E^2\log_2L$ and barriers were IIDs from \EQ{pd_barriers} with variance $\sigma_B^2\log_2L$. Each point on the density plot was averaged over $10^5$ realizations. The condensation phase transition is at $k_BT_{critical} = \sigma _E/\sqrt{2 \ln 2}\approx 0.84 \, \sigma_E$. We notice that the probability of having a direct strong Mpemba effect is finite and even high for certain regions of the $\sigma_E\sigma_B-$ plane for $T_b < T_{critical}$ (left panel).}
\end{figure} 

We also studied the system size dependence of REM with random barriers, see \FIG{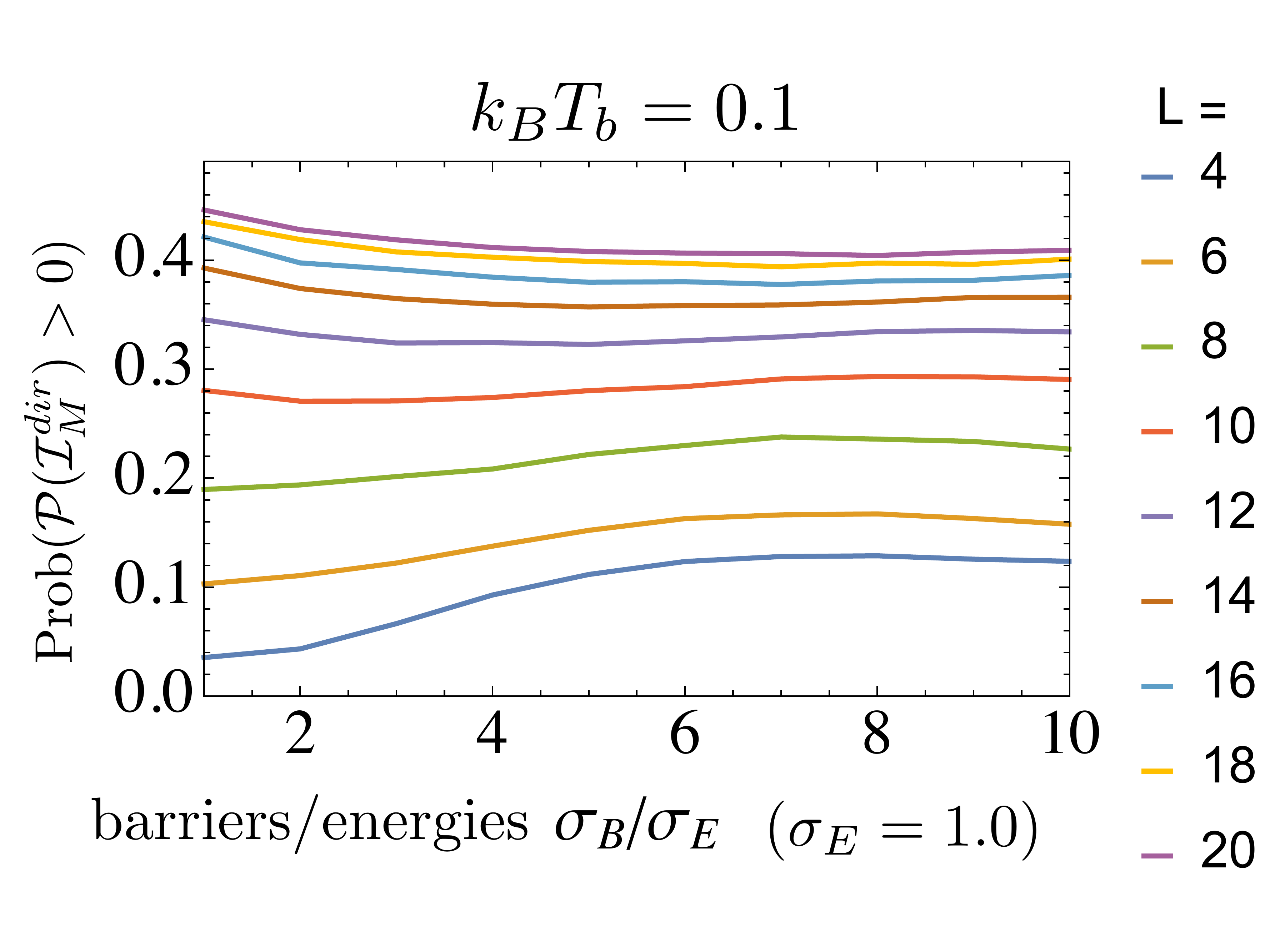}. The system size was varied ($L \in[4,20]$) and we took the bath temperature to be $k_BT_b =0.1$. The energies were IIDs from \EQ{pd_energies} with variance $\sigma_E^2\log_2L$ where $\sigma _E = 1.0$ and barriers were IIDs from \EQ{pd_barriers} with variance $\sigma_B^2\log_2L$. Each point on the density plot was averaged over $2\times 10^5$ realizations. We notice that the  probability of the parity being positive for the direct Mpemba index seems to be converging to a limiting value with increasing system size. Although, we tested small sizes, the convergence suggests the thermodynamic limit behavior. 
\begin{figure}[h]
\includegraphics[width=\linewidth]{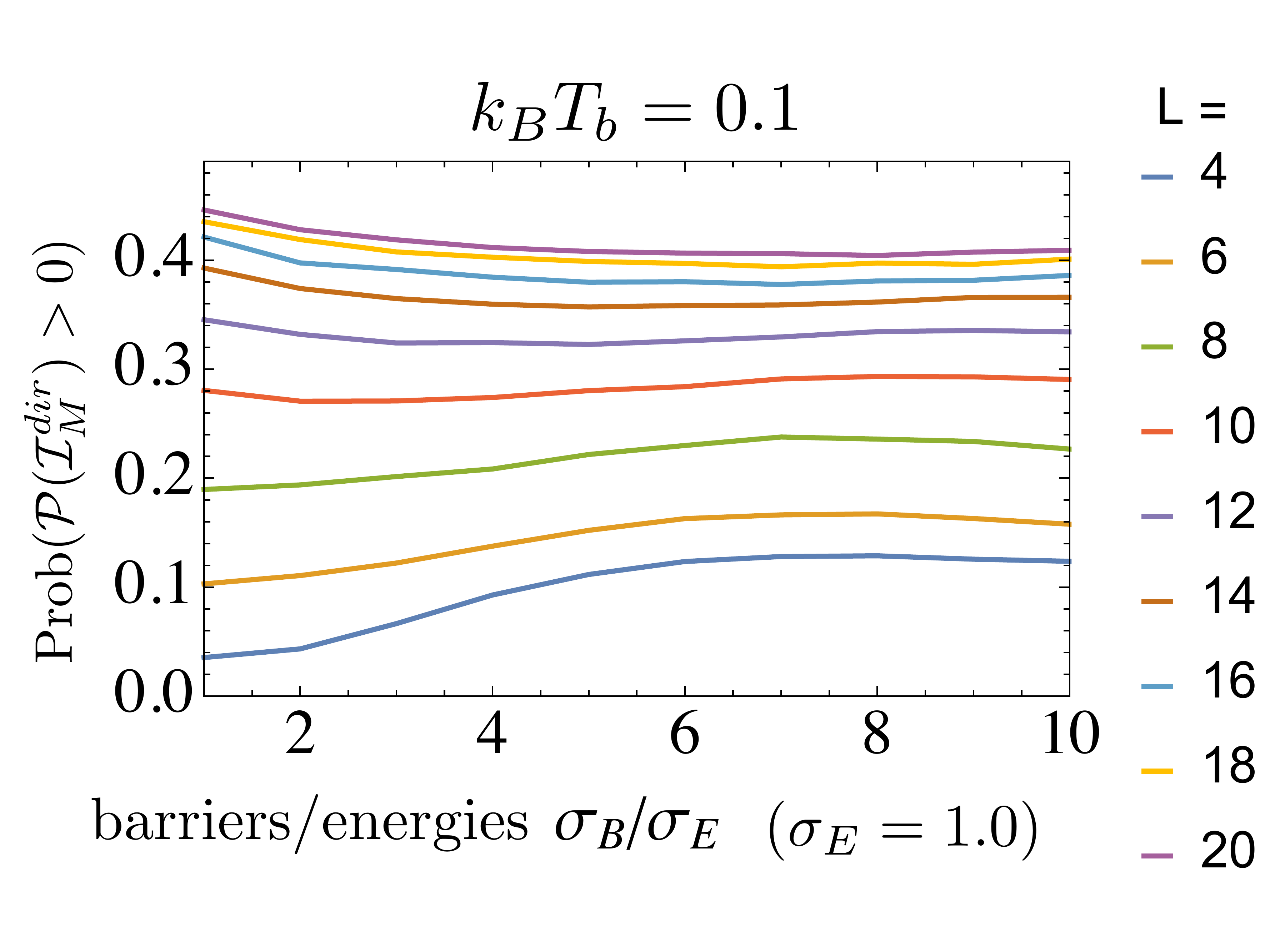}
\caption{\label{fig:fig_11_syssize_Tb01_parity_M20000_v01.pdf} Lower bound for the probability of the strong Mpemba effect -- the probability of the parity being positive for the direct Mpemba index (see \EQ{index_dir}) for the case of REM with random barriers and different system sizes $L\in[4,20]$. The bath temperature is $k_BT_b =0.1$. The energies were IIDs from \EQ{pd_energies} with variance $\sigma_E^2\log_2L$ and $\sigma _E = 1.0$ and barriers were IIDs from \EQ{pd_barriers} with variance $\sigma_B^2\log_2L$. Each point on the density plot was averaged over $2\times 10^5$ realizations. We notice that the probability of the parity being positive for the direct Mpemba index seems to be converging to a finite limiting value with increasing system size.}
\end{figure} 

\section{Discussion}

The Mpemba effect is a "short-cut" in relaxation time. The direct Mpemba effect implies that initiating the system at a particular hot temperature results in cooling down which is faster than any colder temperature, when the system is coupled to a cold bath. Possibly even more counter-intuitive is the inverse Mpemba effect where an analog effect happens in heating. 
Similarly to the direct Mpemba effect, in annealing one first heat the system and then cools it in a controlled manner such that it acquires desirable features (relaxes to the ground state, has fewer defects, etc.). More specifically, simulated annealing is a probabilistic technique used to find ground states~\cite{Khachaturyan:a19748,79Khachaturyan,Kirkpatrick671}, while annealing in metallurgy is used to make materials with larger mono-crystal domains and fewer defects~\cite{Sohn20053}. It would be interesting to explore the connection between annealing and the Mpemba effect. 

Markov Chain Monte Carlo (MCMC) algorithms are essential numerical tools broadly used in many branches of science to estimate steady-state properties of various systems~\cite{97Sokal}. It is often desirable to speed up the relaxation of a MCMC to the steady state, see, e.g., ~\cite{turitsyn2011irreversible, LiftingVucelja}. Our results serve as a proof of principle that in specific systems one could devise additional transition barriers  ($B_{ij}$s) that would cause speed up of a MCMC algorithm's relaxation to equilibrium by creating a strong Mpemba effect. 

Approach to equilibrium often has a non-trivial relationship with the energy landscape and nature of the barriers. This is especially true in glassy materials and complex many-body systems. The approach to equilibration can even be used to explore structures in glassy systems and many-body systems experimentally~\cite{Samarakoon18102016}. One of the future directions is to deepen the understanding of the relation of the Mpemba effect and to the plethora of nontrivial cooling phenomena present in glassy materials; such as memory, aging, and rejuvenation.

{\bf Acknowledgement:} The work of IK was supported by the NSF grant DMR-1508245. This research was supported in part by the National Science Foundation under Grant No. NSF PHY-1125915. MV thanks E.~Siggia, G.~Falkovich, J.~Cohen and C.~Kirst for insightful remarks. OR is supported by a research grant from Mr. and Mrs. Dan Kane  and the Abramson Family Center for Young Scientists. OR thanks C.~Jarzynski for insightful remarks.

\bibliography{MpembaBibNoPaperFiles.bib}

\begin{thebibliography}{40}%
\makeatletter
\providecommand \@ifxundefined [1]{%
 \@ifx{#1\undefined}
}%
\providecommand \@ifnum [1]{%
 \ifnum #1\expandafter \@firstoftwo
 \else \expandafter \@secondoftwo
 \fi
}%
\providecommand \@ifx [1]{%
 \ifx #1\expandafter \@firstoftwo
 \else \expandafter \@secondoftwo
 \fi
}%
\providecommand \natexlab [1]{#1}%
\providecommand \enquote  [1]{``#1''}%
\providecommand \bibnamefont  [1]{#1}%
\providecommand \bibfnamefont [1]{#1}%
\providecommand \citenamefont [1]{#1}%
\providecommand \href@noop [0]{\@secondoftwo}%
\providecommand \href [0]{\begingroup \@sanitize@url \@href}%
\providecommand \@href[1]{\@@startlink{#1}\@@href}%
\providecommand \@@href[1]{\endgroup#1\@@endlink}%
\providecommand \@sanitize@url [0]{\catcode `\\12\catcode `\$12\catcode
  `\&12\catcode `\#12\catcode `\^12\catcode `\_12\catcode `\%12\relax}%
\providecommand \@@startlink[1]{}%
\providecommand \@@endlink[0]{}%
\providecommand \url  [0]{\begingroup\@sanitize@url \@url }%
\providecommand \@url [1]{\endgroup\@href {#1}{\urlprefix }}%
\providecommand \urlprefix  [0]{URL }%
\providecommand \Eprint [0]{\href }%
\providecommand \doibase [0]{http://dx.doi.org/}%
\providecommand \selectlanguage [0]{\@gobble}%
\providecommand \bibinfo  [0]{\@secondoftwo}%
\providecommand \bibfield  [0]{\@secondoftwo}%
\providecommand \translation [1]{[#1]}%
\providecommand \BibitemOpen [0]{}%
\providecommand \bibitemStop [0]{}%
\providecommand \bibitemNoStop [0]{.\EOS\space}%
\providecommand \EOS [0]{\spacefactor3000\relax}%
\providecommand \BibitemShut  [1]{\csname bibitem#1\endcsname}%
\let\auto@bib@innerbib\@empty
\bibitem [{\citenamefont {Aristotle}()}]{Aristotle}%
  \BibitemOpen
  \bibfield  {author} {\bibinfo {author} {\bibnamefont {Aristotle}},\
  }\href@noop {} {\emph {\bibinfo {title} {Meteorologica}}},\ \bibinfo {note}
  {translated by H. D. P. Lee, (Harvard U. P. 1962) Book I, Chap. XII, pp.
  85-87}\BibitemShut {NoStop}%
\bibitem [{\citenamefont {Mpemba}\ and\ \citenamefont
  {Osborne}(1969)}]{Mpemba}%
  \BibitemOpen
  \bibfield  {author} {\bibinfo {author} {\bibfnamefont {E~B}\ \bibnamefont
  {Mpemba}}\ and\ \bibinfo {author} {\bibfnamefont {D~G}\ \bibnamefont
  {Osborne}},\ }\bibfield  {title} {\enquote {\bibinfo {title} {Cool?}}\ }\href
  {http://stacks.iop.org/0031-9120/4/i=3/a=312} {\bibfield  {journal} {\bibinfo
   {journal} {Physics Education}\ }\textbf {\bibinfo {volume} {4}},\ \bibinfo
  {pages} {172} (\bibinfo {year} {1969})}\BibitemShut {NoStop}%
\bibitem [{\citenamefont {Mirabedin}\ and\ \citenamefont
  {Farhadi}(2017)}]{mirabedin2017numerical}%
  \BibitemOpen
  \bibfield  {author} {\bibinfo {author} {\bibfnamefont {Seyed~Milad}\
  \bibnamefont {Mirabedin}}\ and\ \bibinfo {author} {\bibfnamefont {Fatola}\
  \bibnamefont {Farhadi}},\ }\bibfield  {title} {\enquote {\bibinfo {title}
  {Numerical investigation of solidification of single droplets with and
  without evaporation mechanism},}\ }\href@noop {} {\bibfield  {journal}
  {\bibinfo  {journal} {International Journal of Refrigeration}\ }\textbf
  {\bibinfo {volume} {73}},\ \bibinfo {pages} {219--225} (\bibinfo {year}
  {2017})}\BibitemShut {NoStop}%
\bibitem [{\citenamefont {Auerbach}(1995)}]{auerbach1995supercooling}%
  \BibitemOpen
  \bibfield  {author} {\bibinfo {author} {\bibfnamefont {David}\ \bibnamefont
  {Auerbach}},\ }\bibfield  {title} {\enquote {\bibinfo {title} {Supercooling
  and the mpemba effect: When hot water freezes quicker than cold},}\
  }\href@noop {} {\bibfield  {journal} {\bibinfo  {journal} {American Journal
  of Physics}\ }\textbf {\bibinfo {volume} {63}},\ \bibinfo {pages} {882--885}
  (\bibinfo {year} {1995})}\BibitemShut {NoStop}%
\bibitem [{\citenamefont {Vynnycky}\ and\ \citenamefont
  {Kimura}(2015)}]{Vynnycky2015243}%
  \BibitemOpen
  \bibfield  {author} {\bibinfo {author} {\bibfnamefont {M.}~\bibnamefont
  {Vynnycky}}\ and\ \bibinfo {author} {\bibfnamefont {S.}~\bibnamefont
  {Kimura}},\ }\bibfield  {title} {\enquote {\bibinfo {title} {Can natural
  convection alone explain the mpemba effect?}}\ }\href {\doibase
  https://doi.org/10.1016/j.ijheatmasstransfer.2014.09.015} {\bibfield
  {journal} {\bibinfo  {journal} {International Journal of Heat and Mass
  Transfer}\ }\textbf {\bibinfo {volume} {80}},\ \bibinfo {pages} {243 -- 255}
  (\bibinfo {year} {2015})}\BibitemShut {NoStop}%
\bibitem [{\citenamefont {Zhang}\ \emph {et~al.}(2014)\citenamefont {Zhang},
  \citenamefont {Huang}, \citenamefont {Ma}, \citenamefont {Zhou},
  \citenamefont {Zhou}, \citenamefont {Zheng}, \citenamefont {Jiang},\ and\
  \citenamefont {Sun}}]{HydroBond_zhang2014hydrogen}%
  \BibitemOpen
  \bibfield  {author} {\bibinfo {author} {\bibfnamefont {Xi}~\bibnamefont
  {Zhang}}, \bibinfo {author} {\bibfnamefont {Yongli}\ \bibnamefont {Huang}},
  \bibinfo {author} {\bibfnamefont {Zengsheng}\ \bibnamefont {Ma}}, \bibinfo
  {author} {\bibfnamefont {Yichun}\ \bibnamefont {Zhou}}, \bibinfo {author}
  {\bibfnamefont {Ji}~\bibnamefont {Zhou}}, \bibinfo {author} {\bibfnamefont
  {Weitao}\ \bibnamefont {Zheng}}, \bibinfo {author} {\bibfnamefont {Qing}\
  \bibnamefont {Jiang}}, \ and\ \bibinfo {author} {\bibfnamefont {Chang~Q}\
  \bibnamefont {Sun}},\ }\bibfield  {title} {\enquote {\bibinfo {title}
  {Hydrogen-bond memory and water-skin supersolidity resolving the mpemba
  paradox},}\ }\href@noop {} {\bibfield  {journal} {\bibinfo  {journal}
  {Physical Chemistry Chemical Physics}\ }\textbf {\bibinfo {volume} {16}},\
  \bibinfo {pages} {22995--23002} (\bibinfo {year} {2014})}\BibitemShut
  {NoStop}%
\bibitem [{\citenamefont {Tao}\ \emph {et~al.}(2016)\citenamefont {Tao},
  \citenamefont {Zou}, \citenamefont {Jia}, \citenamefont {Li},\ and\
  \citenamefont {Cremer}}]{DFT_Numerical_Mpemba}%
  \BibitemOpen
  \bibfield  {author} {\bibinfo {author} {\bibfnamefont {Yunwen}\ \bibnamefont
  {Tao}}, \bibinfo {author} {\bibfnamefont {Wenli}\ \bibnamefont {Zou}},
  \bibinfo {author} {\bibfnamefont {Junteng}\ \bibnamefont {Jia}}, \bibinfo
  {author} {\bibfnamefont {Wei}\ \bibnamefont {Li}}, \ and\ \bibinfo {author}
  {\bibfnamefont {Dieter}\ \bibnamefont {Cremer}},\ }\bibfield  {title}
  {\enquote {\bibinfo {title} {Different ways of hydrogen bonding in water-why
  does warm water freeze faster than cold water?}}\ }\href@noop {} {\bibfield
  {journal} {\bibinfo  {journal} {Journal of chemical theory and computation}\
  }\textbf {\bibinfo {volume} {13}},\ \bibinfo {pages} {55--76} (\bibinfo
  {year} {2016})}\BibitemShut {NoStop}%
\bibitem [{\citenamefont {Katz}(2009)}]{Katz}%
  \BibitemOpen
  \bibfield  {author} {\bibinfo {author} {\bibfnamefont {J.~I.}\ \bibnamefont
  {Katz}},\ }\bibfield  {title} {\enquote {\bibinfo {title} {When hot water
  freezes before cold},}\ }\href@noop {} {\bibfield  {journal} {\bibinfo
  {journal} {American Journal of Physics}\ }\textbf {\bibinfo {volume} {77}},\
  \bibinfo {pages} {27--29} (\bibinfo {year} {2009})}\BibitemShut {NoStop}%
\bibitem [{\citenamefont {Brownridge}(2011)}]{brownridge2011does}%
  \BibitemOpen
  \bibfield  {author} {\bibinfo {author} {\bibfnamefont {James~D}\ \bibnamefont
  {Brownridge}},\ }\bibfield  {title} {\enquote {\bibinfo {title} {When does
  hot water freeze faster then cold water? a search for the mpemba effect},}\
  }\href@noop {} {\bibfield  {journal} {\bibinfo  {journal} {American Journal
  of Physics}\ }\textbf {\bibinfo {volume} {79}},\ \bibinfo {pages} {78--84}
  (\bibinfo {year} {2011})}\BibitemShut {NoStop}%
\bibitem [{\citenamefont {Burridge}\ and\ \citenamefont
  {Linden}(2016)}]{burridge2016questioning}%
  \BibitemOpen
  \bibfield  {author} {\bibinfo {author} {\bibfnamefont {Henry~C}\ \bibnamefont
  {Burridge}}\ and\ \bibinfo {author} {\bibfnamefont {Paul~F}\ \bibnamefont
  {Linden}},\ }\bibfield  {title} {\enquote {\bibinfo {title} {Questioning the
  mpemba effect: hot water does not cool more quickly than cold},}\ }\href@noop
  {} {\bibfield  {journal} {\bibinfo  {journal} {Scientific Reports}\ }\textbf
  {\bibinfo {volume} {6}} (\bibinfo {year} {2016})}\BibitemShut {NoStop}%
\bibitem [{\citenamefont {Katz}(2017)}]{katz2017reply}%
  \BibitemOpen
  \bibfield  {author} {\bibinfo {author} {\bibfnamefont {Jonathan~I}\
  \bibnamefont {Katz}},\ }\bibfield  {title} {\enquote {\bibinfo {title} {Reply
  to burridge \& linden: Hot water may freeze sooner than cold},}\ }\href@noop
  {} {\bibfield  {journal} {\bibinfo  {journal} {arXiv preprint
  arXiv:1701.03219}\ } (\bibinfo {year} {2017})}\BibitemShut {NoStop}%
\bibitem [{\citenamefont {Chaddah}\ \emph {et~al.}(2010)\citenamefont
  {Chaddah}, \citenamefont {Dash}, \citenamefont {Kumar},\ and\ \citenamefont
  {Banerjee}}]{Magnetic_Mpemba}%
  \BibitemOpen
  \bibfield  {author} {\bibinfo {author} {\bibfnamefont {P}~\bibnamefont
  {Chaddah}}, \bibinfo {author} {\bibfnamefont {S}~\bibnamefont {Dash}},
  \bibinfo {author} {\bibfnamefont {Kranti}\ \bibnamefont {Kumar}}, \ and\
  \bibinfo {author} {\bibfnamefont {A}~\bibnamefont {Banerjee}},\ }\bibfield
  {title} {\enquote {\bibinfo {title} {Overtaking while approaching
  equilibrium},}\ }\href@noop {} {\bibfield  {journal} {\bibinfo  {journal}
  {arXiv preprint arXiv:1011.3598}\ } (\bibinfo {year} {2010})}\BibitemShut
  {NoStop}%
\bibitem [{\citenamefont {Greaney}\ \emph {et~al.}(2011)\citenamefont
  {Greaney}, \citenamefont {Lani}, \citenamefont {Cicero},\ and\ \citenamefont
  {Grossman}}]{Theory_Carbon_nano_greaney2011mpemba}%
  \BibitemOpen
  \bibfield  {author} {\bibinfo {author} {\bibfnamefont {P~Alex}\ \bibnamefont
  {Greaney}}, \bibinfo {author} {\bibfnamefont {Giovanna}\ \bibnamefont
  {Lani}}, \bibinfo {author} {\bibfnamefont {Giancarlo}\ \bibnamefont
  {Cicero}}, \ and\ \bibinfo {author} {\bibfnamefont {Jeffrey~C}\ \bibnamefont
  {Grossman}},\ }\bibfield  {title} {\enquote {\bibinfo {title} {Mpemba-like
  behavior in carbon nanotube resonators},}\ }\href@noop {} {\bibfield
  {journal} {\bibinfo  {journal} {Metallurgical and Materials Transactions A}\
  }\textbf {\bibinfo {volume} {42}},\ \bibinfo {pages} {3907--3912} (\bibinfo
  {year} {2011})}\BibitemShut {NoStop}%
\bibitem [{\citenamefont {Lasanta}\ \emph {et~al.}(2017)\citenamefont
  {Lasanta}, \citenamefont {Vega~Reyes}, \citenamefont {Prados},\ and\
  \citenamefont {Santos}}]{Granular_Mpemba_PhysRevLett}%
  \BibitemOpen
  \bibfield  {author} {\bibinfo {author} {\bibfnamefont {Antonio}\ \bibnamefont
  {Lasanta}}, \bibinfo {author} {\bibfnamefont {Francisco}\ \bibnamefont
  {Vega~Reyes}}, \bibinfo {author} {\bibfnamefont {Antonio}\ \bibnamefont
  {Prados}}, \ and\ \bibinfo {author} {\bibfnamefont {Andr\'es}\ \bibnamefont
  {Santos}},\ }\bibfield  {title} {\enquote {\bibinfo {title} {When the hotter
  cools more quickly: Mpemba effect in granular fluids},}\ }\href {\doibase
  10.1103/PhysRevLett.119.148001} {\bibfield  {journal} {\bibinfo  {journal}
  {Phys. Rev. Lett.}\ }\textbf {\bibinfo {volume} {119}},\ \bibinfo {pages}
  {148001} (\bibinfo {year} {2017})}\BibitemShut {NoStop}%
\bibitem [{\citenamefont {Ahn}\ \emph {et~al.}(2016)\citenamefont {Ahn},
  \citenamefont {Kang}, \citenamefont {Koh},\ and\ \citenamefont
  {Lee}}]{paper:hydrates}%
  \BibitemOpen
  \bibfield  {author} {\bibinfo {author} {\bibfnamefont {Yun-Ho}\ \bibnamefont
  {Ahn}}, \bibinfo {author} {\bibfnamefont {Hyery}\ \bibnamefont {Kang}},
  \bibinfo {author} {\bibfnamefont {Dong-Yeun}\ \bibnamefont {Koh}}, \ and\
  \bibinfo {author} {\bibfnamefont {Huen}\ \bibnamefont {Lee}},\ }\bibfield
  {title} {\enquote {\bibinfo {title} {Experimental verifications of
  mpemba-like behaviors of clathrate hydrates},}\ }\href@noop {} {\bibfield
  {journal} {\bibinfo  {journal} {Korean Journal of Chemical Engineering}\ ,\
  \bibinfo {pages} {1--5}} (\bibinfo {year} {2016})}\BibitemShut {NoStop}%
\bibitem [{\citenamefont {Hu}\ \emph {et~al.}(2018)\citenamefont {Hu},
  \citenamefont {Li}, \citenamefont {Huang}, \citenamefont {Li}, \citenamefont
  {Luo}, \citenamefont {Chen}, \citenamefont {Jiang},\ and\ \citenamefont
  {An}}]{18Mpembapoly}%
  \BibitemOpen
  \bibfield  {author} {\bibinfo {author} {\bibfnamefont {Cunliang}\
  \bibnamefont {Hu}}, \bibinfo {author} {\bibfnamefont {Jingqing}\ \bibnamefont
  {Li}}, \bibinfo {author} {\bibfnamefont {Shaoyong}\ \bibnamefont {Huang}},
  \bibinfo {author} {\bibfnamefont {Hongfei}\ \bibnamefont {Li}}, \bibinfo
  {author} {\bibfnamefont {Chuanfu}\ \bibnamefont {Luo}}, \bibinfo {author}
  {\bibfnamefont {Jizhong}\ \bibnamefont {Chen}}, \bibinfo {author}
  {\bibfnamefont {Shichun}\ \bibnamefont {Jiang}}, \ and\ \bibinfo {author}
  {\bibfnamefont {Lijia}\ \bibnamefont {An}},\ }\bibfield  {title} {\enquote
  {\bibinfo {title} {Conformation directed mpemba effect on polylactide
  crystallization},}\ }\href@noop {} {\bibfield  {journal} {\bibinfo  {journal}
  {Crystal Growth \& Design}\ }\textbf {\bibinfo {volume} {18}},\ \bibinfo
  {pages} {5757--5762} (\bibinfo {year} {2018})}\BibitemShut {NoStop}%
\bibitem [{\citenamefont {Keller}\ \emph {et~al.}(2018)\citenamefont {Keller},
  \citenamefont {Torggler}, \citenamefont {J{\"a}ger}, \citenamefont
  {Sch{\"u}tz}, \citenamefont {Ritsch},\ and\ \citenamefont
  {Morigi}}]{keller2018quenches}%
  \BibitemOpen
  \bibfield  {author} {\bibinfo {author} {\bibfnamefont {Tim}\ \bibnamefont
  {Keller}}, \bibinfo {author} {\bibfnamefont {Valentin}\ \bibnamefont
  {Torggler}}, \bibinfo {author} {\bibfnamefont {Simon~B}\ \bibnamefont
  {J{\"a}ger}}, \bibinfo {author} {\bibfnamefont {Stefan}\ \bibnamefont
  {Sch{\"u}tz}}, \bibinfo {author} {\bibfnamefont {Helmut}\ \bibnamefont
  {Ritsch}}, \ and\ \bibinfo {author} {\bibfnamefont {Giovanna}\ \bibnamefont
  {Morigi}},\ }\bibfield  {title} {\enquote {\bibinfo {title} {Quenches across
  the self-organization transition in multimode cavities},}\ }\href@noop {}
  {\bibfield  {journal} {\bibinfo  {journal} {New Journal of Physics}\ }\textbf
  {\bibinfo {volume} {20}},\ \bibinfo {pages} {025004} (\bibinfo {year}
  {2018})}\BibitemShut {NoStop}%
\bibitem [{\citenamefont {Lu}\ and\ \citenamefont {Raz}(2017)}]{2016LuRaz}%
  \BibitemOpen
  \bibfield  {author} {\bibinfo {author} {\bibfnamefont {Zhiyue}\ \bibnamefont
  {Lu}}\ and\ \bibinfo {author} {\bibfnamefont {Oren}\ \bibnamefont {Raz}},\
  }\bibfield  {title} {\enquote {\bibinfo {title} {Nonequilibrium
  thermodynamics of the markovian mpemba effect and its inverse},}\ }\href@noop
  {} {\bibfield  {journal} {\bibinfo  {journal} {Proceedings of the National
  Academy of Sciences}\ }\textbf {\bibinfo {volume} {114}},\ \bibinfo {pages}
  {5083--5088} (\bibinfo {year} {2017})}\BibitemShut {NoStop}%
\bibitem [{\citenamefont {van Kampen N.~G.}(2001)}]{van_Kampen}%
  \BibitemOpen
  \bibfield  {author} {\bibinfo {author} {\bibnamefont {van Kampen N.~G.}},\
  }\href@noop {} {\emph {\bibinfo {title} {Stochastic processes in physics and
  chemistry}}}\ (\bibinfo  {publisher} {Elsevier},\ \bibinfo {year}
  {2001})\BibitemShut {NoStop}%
\bibitem [{Note1()}]{Note1}%
  \BibitemOpen
  \bibinfo {note} {For simplicity, we consider here only ergodic finite state
  systems. Much of the analysis can be easily generalized to infinite systems
  as well.}\BibitemShut {Stop}%
\bibitem [{\citenamefont {Mandal}\ and\ \citenamefont
  {Jarzynski}(2011)}]{2011MandalJarzynski}%
  \BibitemOpen
  \bibfield  {author} {\bibinfo {author} {\bibfnamefont {Dibyendu}\
  \bibnamefont {Mandal}}\ and\ \bibinfo {author} {\bibfnamefont {Christopher}\
  \bibnamefont {Jarzynski}},\ }\bibfield  {title} {\enquote {\bibinfo {title}
  {A proof by graphical construction of the no-pumping theorem of stochastic
  pumps},}\ }\href {http://stacks.iop.org/1742-5468/2011/i=10/a=P10006}
  {\bibfield  {journal} {\bibinfo  {journal} {Journal of Statistical Mechanics:
  Theory and Experiment}\ }\textbf {\bibinfo {volume} {2011}},\ \bibinfo
  {pages} {P10006} (\bibinfo {year} {2011})}\BibitemShut {NoStop}%
\bibitem [{Note2()}]{Note2}%
  \BibitemOpen
  \bibinfo {note} {For detailed balance matrices $R$, the eigenvalues are in
  fact all real. We use this more general notation as our discussion can also
  be relevant to $R$'s that do not satisfy detailed balance.}\BibitemShut
  {Stop}%
\bibitem [{Note3()}]{Note3}%
  \BibitemOpen
  \bibinfo {note} {The above definition to the Mpemba effect is readily
  generalizable for the degenerate case, $|\Re \lambda _2| = |\Re \lambda
  _3|$.}\BibitemShut {Stop}%
\bibitem [{Note4()}]{Note4}%
  \BibitemOpen
  \bibinfo {note} {Since the relaxation described by Eq. \protect \textup
  {\hbox {\mathsurround \z@ \protect \normalfont (\ignorespaces \ref
  {eq:MasterEq}\unskip \@@italiccorr )}} reaches the bath's Boltzmann
  distribution at infinite times, the actual observed relaxation time may
  depend on the choice of a distance function on the probability simplex. The
  distance function may be relative entropy or other measures, as long as the
  exponential ratio between the time dependent coefficients of ${v_3}$ and
  $v_2$ is not compensated by the fact that the distances are measured along
  different directions. In other words, the metric does not grow exponentially
  faster in one direction compared to the other.}\BibitemShut {Stop}%
\bibitem [{\citenamefont {Vives}\ \emph {et~al.}(1997)\citenamefont {Vives},
  \citenamefont {Cast{\'a}n},\ and\ \citenamefont
  {Planes}}]{Mean_field_MJP_1997unified}%
  \BibitemOpen
  \bibfield  {author} {\bibinfo {author} {\bibfnamefont {Eduard}\ \bibnamefont
  {Vives}}, \bibinfo {author} {\bibfnamefont {Teresa}\ \bibnamefont
  {Cast{\'a}n}}, \ and\ \bibinfo {author} {\bibfnamefont {Antoni}\ \bibnamefont
  {Planes}},\ }\bibfield  {title} {\enquote {\bibinfo {title} {Unified
  mean-field study of ferro-and antiferromagnetic behavior of the ising model
  with external field},}\ }\href@noop {} {\bibfield  {journal} {\bibinfo
  {journal} {American Journal of Physics}\ }\textbf {\bibinfo {volume} {65}},\
  \bibinfo {pages} {907--913} (\bibinfo {year} {1997})}\BibitemShut {NoStop}%
\bibitem [{\citenamefont {Glauber}(1963)}]{Glauber}%
  \BibitemOpen
  \bibfield  {author} {\bibinfo {author} {\bibfnamefont {Roy~J.}\ \bibnamefont
  {Glauber}},\ }\bibfield  {title} {\enquote {\bibinfo {title} {Time dependent
  statistics of the ising model},}\ }\href@noop {} {\bibfield  {journal}
  {\bibinfo  {journal} {J. Math. Phys.}\ }\textbf {\bibinfo {volume} {4}},\
  \bibinfo {pages} {294--307} (\bibinfo {year} {1963})}\BibitemShut {NoStop}%
\bibitem [{\citenamefont {Lapas}\ \emph {et~al.}(2015)\citenamefont {Lapas},
  \citenamefont {Ferreira}, \citenamefont {Rub{\'\i}},\ and\ \citenamefont
  {Oliveira}}]{anomalous2015}%
  \BibitemOpen
  \bibfield  {author} {\bibinfo {author} {\bibfnamefont {Luciano~C}\
  \bibnamefont {Lapas}}, \bibinfo {author} {\bibfnamefont {Rogelma~MS}\
  \bibnamefont {Ferreira}}, \bibinfo {author} {\bibfnamefont {J~Miguel}\
  \bibnamefont {Rub{\'\i}}}, \ and\ \bibinfo {author} {\bibfnamefont
  {Fernando~A}\ \bibnamefont {Oliveira}},\ }\bibfield  {title} {\enquote
  {\bibinfo {title} {Anomalous law of cooling},}\ }\href@noop {} {\bibfield
  {journal} {\bibinfo  {journal} {The Journal of chemical physics}\ }\textbf
  {\bibinfo {volume} {142}},\ \bibinfo {pages} {104106} (\bibinfo {year}
  {2015})}\BibitemShut {NoStop}%
\bibitem [{\citenamefont {Mu{\~n}oz-Tapia}\ \emph {et~al.}(2017)\citenamefont
  {Mu{\~n}oz-Tapia}, \citenamefont {Brito},\ and\ \citenamefont
  {Parrondo}}]{Parrondo2017heating}%
  \BibitemOpen
  \bibfield  {author} {\bibinfo {author} {\bibfnamefont {Ram{\'o}n}\
  \bibnamefont {Mu{\~n}oz-Tapia}}, \bibinfo {author} {\bibfnamefont {Ricardo}\
  \bibnamefont {Brito}}, \ and\ \bibinfo {author} {\bibfnamefont {Juan~MR}\
  \bibnamefont {Parrondo}},\ }\bibfield  {title} {\enquote {\bibinfo {title}
  {Heating without heat: thermodynamics of passive energy filters between
  finite systems},}\ }\href@noop {} {\bibfield  {journal} {\bibinfo  {journal}
  {arXiv preprint arXiv:1705.04657}\ } (\bibinfo {year} {2017})}\BibitemShut
  {NoStop}%
\bibitem [{\citenamefont {Derrida}(1980)}]{DERRIDA198029}%
  \BibitemOpen
  \bibfield  {author} {\bibinfo {author} {\bibfnamefont {B.}~\bibnamefont
  {Derrida}},\ }\bibfield  {title} {\enquote {\bibinfo {title} {The random
  energy model},}\ }\href {\doibase
  http://dx.doi.org/10.1016/0370-1573(80)90076-9} {\bibfield  {journal}
  {\bibinfo  {journal} {Physics Reports}\ }\textbf {\bibinfo {volume} {67}},\
  \bibinfo {pages} {29 -- 35} (\bibinfo {year} {1980})}\BibitemShut {NoStop}%
\bibitem [{\citenamefont {{Ben Arous}}\ \emph {et~al.}(2002)\citenamefont {{Ben
  Arous}}, \citenamefont {{Bovier}},\ and\ \citenamefont
  {{Gayrard}}}]{AgingREM2002}%
  \BibitemOpen
  \bibfield  {author} {\bibinfo {author} {\bibfnamefont {G.}~\bibnamefont {{Ben
  Arous}}}, \bibinfo {author} {\bibfnamefont {A.}~\bibnamefont {{Bovier}}}, \
  and\ \bibinfo {author} {\bibfnamefont {V.}~\bibnamefont {{Gayrard}}},\
  }\bibfield  {title} {\enquote {\bibinfo {title} {{Aging in the Random Energy
  Model}},}\ }\href@noop {} {\bibfield  {journal} {\bibinfo  {journal}
  {Physical Review Letters}\ }\textbf {\bibinfo {volume} {88}},\ \bibinfo
  {pages} {087201} (\bibinfo {year} {2002})},\ \Eprint
  {http://arxiv.org/abs/cond-mat/0110223} {cond-mat/0110223} \BibitemShut
  {NoStop}%
\bibitem [{\citenamefont {{Ben Arous}}\ and\ \citenamefont
  {{\v{C}ern\'y}}(2006)}]{BenArousLesHouches}%
  \BibitemOpen
  \bibfield  {author} {\bibinfo {author} {\bibfnamefont {G\'erard}\
  \bibnamefont {{Ben Arous}}}\ and\ \bibinfo {author} {\bibfnamefont
  {Ji\v{r}\'{\i}}\ \bibnamefont {{\v{C}ern\'y}}},\ }\bibfield  {title}
  {\enquote {\bibinfo {title} {Dynamics of trap models},}\ }in\ \href {\doibase
  https://doi.org/10.1016/S0924-8099(06)80045-4} {\emph {\bibinfo {booktitle}
  {Mathematical Statistical Physics}}},\ \bibinfo {series} {Les Houches},
  Vol.~\bibinfo {volume} {83},\ \bibinfo {editor} {edited by\ \bibinfo {editor}
  {\bibfnamefont {Anton}\ \bibnamefont {{Bovier}}}, \bibinfo {editor}
  {\bibfnamefont {Fran\c{c}ois}\ \bibnamefont {{Dunlop}}}, \bibinfo {editor}
  {\bibfnamefont {Aernout}\ \bibnamefont {{van Enter}}}, \bibinfo {editor}
  {\bibfnamefont {Frank}\ \bibnamefont {{den Hollander}}}, \ and\ \bibinfo
  {editor} {\bibfnamefont {Jean}\ \bibnamefont {{Dalibard}}}}\ (\bibinfo
  {publisher} {Elsevier},\ \bibinfo {year} {2006})\ pp.\ \bibinfo {pages}
  {331--394}\BibitemShut {NoStop}%
\bibitem [{\citenamefont {{Baity-Jesi}}\ \emph {et~al.}(2017)\citenamefont
  {{Baity-Jesi}}, \citenamefont {{Biroli}},\ and\ \citenamefont
  {{Cammarota}}}]{TrapRemDynamics2017}%
  \BibitemOpen
  \bibfield  {author} {\bibinfo {author} {\bibfnamefont {Marco}\ \bibnamefont
  {{Baity-Jesi}}}, \bibinfo {author} {\bibfnamefont {Giulio}\ \bibnamefont
  {{Biroli}}}, \ and\ \bibinfo {author} {\bibfnamefont {Chiara}\ \bibnamefont
  {{Cammarota}}},\ }\bibfield  {title} {\enquote {\bibinfo {title} {Activated
  aging dynamics and effective trap model description in the random energy
  model},}\ }\href@noop {} {\bibfield  {journal} {\bibinfo  {journal} {arXiv
  preprint arXiv:1708.03268}\ } (\bibinfo {year} {2017})}\BibitemShut {NoStop}%
\bibitem [{\citenamefont {Khachaturyan}\ \emph {et~al.}(1981)\citenamefont
  {Khachaturyan}, \citenamefont {Semenovsovskaya},\ and\ \citenamefont
  {Vainshtein}}]{Khachaturyan:a19748}%
  \BibitemOpen
  \bibfield  {author} {\bibinfo {author} {\bibfnamefont {A.}~\bibnamefont
  {Khachaturyan}}, \bibinfo {author} {\bibfnamefont {S.}~\bibnamefont
  {Semenovsovskaya}}, \ and\ \bibinfo {author} {\bibfnamefont {B.}~\bibnamefont
  {Vainshtein}},\ }\bibfield  {title} {\enquote {\bibinfo {title} {{The
  thermodynamic approach to the structure analysis of crystals}},}\ }\href@noop
  {} {\bibfield  {journal} {\bibinfo  {journal} {Acta Crystallographica Section
  A}\ }\textbf {\bibinfo {volume} {37}},\ \bibinfo {pages} {742--754} (\bibinfo
  {year} {1981})}\BibitemShut {NoStop}%
\bibitem [{\citenamefont {Khachaturyan}\ \emph {et~al.}(1979)\citenamefont
  {Khachaturyan}, \citenamefont {Semenovsovskaya},\ and\ \citenamefont
  {Vainshtein}}]{79Khachaturyan}%
  \BibitemOpen
  \bibfield  {author} {\bibinfo {author} {\bibfnamefont {A.}~\bibnamefont
  {Khachaturyan}}, \bibinfo {author} {\bibfnamefont {S.}~\bibnamefont
  {Semenovsovskaya}}, \ and\ \bibinfo {author} {\bibfnamefont {B.}~\bibnamefont
  {Vainshtein}},\ }\bibfield  {title} {\enquote {\bibinfo {title}
  {{Statistical-Thermodynamic Approach to Determination of Structure Amplitude
  Phases}},}\ }\href@noop {} {\bibfield  {journal} {\bibinfo  {journal} {Soy.
  Phys. Crystallogr.}\ }\textbf {\bibinfo {volume} {24}},\ \bibinfo {pages}
  {519--524} (\bibinfo {year} {1979})}\BibitemShut {NoStop}%
\bibitem [{\citenamefont {Kirkpatrick}\ \emph {et~al.}(1983)\citenamefont
  {Kirkpatrick}, \citenamefont {Gelatt},\ and\ \citenamefont
  {Vecchi}}]{Kirkpatrick671}%
  \BibitemOpen
  \bibfield  {author} {\bibinfo {author} {\bibfnamefont {S.}~\bibnamefont
  {Kirkpatrick}}, \bibinfo {author} {\bibfnamefont {C.~D.}\ \bibnamefont
  {Gelatt}}, \ and\ \bibinfo {author} {\bibfnamefont {M.~P.}\ \bibnamefont
  {Vecchi}},\ }\bibfield  {title} {\enquote {\bibinfo {title} {Optimization by
  simulated annealing},}\ }\href@noop {} {\bibfield  {journal} {\bibinfo
  {journal} {Science}\ }\textbf {\bibinfo {volume} {220}},\ \bibinfo {pages}
  {671--680} (\bibinfo {year} {1983})}\BibitemShut {NoStop}%
\bibitem [{\citenamefont {Sohn}\ and\ \citenamefont
  {Sridhar}(2005)}]{Sohn20053}%
  \BibitemOpen
  \bibfield  {author} {\bibinfo {author} {\bibfnamefont {H.Y.}\ \bibnamefont
  {Sohn}}\ and\ \bibinfo {author} {\bibfnamefont {S.}~\bibnamefont {Sridhar}},\
  }\bibfield  {title} {\enquote {\bibinfo {title} {1 - descriptions of
  high-temperature metallurgical processes},}\ }in\ \href@noop {} {\emph
  {\bibinfo {booktitle} {Fundamentals of Metallurgy}}},\ \bibinfo {series and
  number} {Woodhead Publishing Series in Metals and Surface Engineering},\
  \bibinfo {editor} {edited by\ \bibinfo {editor} {\bibfnamefont {Seshadri}\
  \bibnamefont {Seetharaman}}}\ (\bibinfo  {publisher} {Woodhead Publishing},\
  \bibinfo {year} {2005})\ pp.\ \bibinfo {pages} {3 -- 37}\BibitemShut
  {NoStop}%
\bibitem [{\citenamefont {Sokal}(1997)}]{97Sokal}%
  \BibitemOpen
  \bibfield  {author} {\bibinfo {author} {\bibfnamefont {A.}~\bibnamefont
  {Sokal}},\ }\bibfield  {title} {\enquote {\bibinfo {title} {Monte carlo
  methods in statistical mechanics: Foundations and new algorithms},}\ }in\
  \href@noop {} {\emph {\bibinfo {booktitle} {Functional Integration}}},\
  \bibinfo {series} {NATO ASI Series}, Vol.\ \bibinfo {volume} {361},\ \bibinfo
  {editor} {edited by\ \bibinfo {editor} {\bibfnamefont {Cecile}\ \bibnamefont
  {DeWitt-Morette}}, \bibinfo {editor} {\bibfnamefont {Pierre}\ \bibnamefont
  {Cartier}}, \ and\ \bibinfo {editor} {\bibfnamefont {Antoine}\ \bibnamefont
  {Folacci}}}\ (\bibinfo  {publisher} {Springer US},\ \bibinfo {address} {New
  York, NY, USA},\ \bibinfo {year} {1997})\ pp.\ \bibinfo {pages}
  {131--192}\BibitemShut {NoStop}%
\bibitem [{\citenamefont {Turitsyn}\ \emph {et~al.}(2011)\citenamefont
  {Turitsyn}, \citenamefont {Chertkov},\ and\ \citenamefont
  {Vucelja}}]{turitsyn2011irreversible}%
  \BibitemOpen
  \bibfield  {author} {\bibinfo {author} {\bibfnamefont {K.S.}\ \bibnamefont
  {Turitsyn}}, \bibinfo {author} {\bibfnamefont {M.}~\bibnamefont {Chertkov}},
  \ and\ \bibinfo {author} {\bibfnamefont {M.}~\bibnamefont {Vucelja}},\
  }\bibfield  {title} {\enquote {\bibinfo {title} {Irreversible monte carlo
  algorithms for efficient sampling},}\ }\href@noop {} {\bibfield  {journal}
  {\bibinfo  {journal} {Physica D Nonlinear Phenomena}\ }\textbf {\bibinfo
  {volume} {240}},\ \bibinfo {pages} {410--414} (\bibinfo {year}
  {2011})}\BibitemShut {NoStop}%
\bibitem [{\citenamefont {Vucelja}(2016)}]{LiftingVucelja}%
  \BibitemOpen
  \bibfield  {author} {\bibinfo {author} {\bibfnamefont {Marija}\ \bibnamefont
  {Vucelja}},\ }\bibfield  {title} {\enquote {\bibinfo {title} {{Lifting--A
  nonreversible Markov chain Monte Carlo algorithm}},}\ }\href@noop {}
  {\bibfield  {journal} {\bibinfo  {journal} {American Journal of Physics}\
  }\textbf {\bibinfo {volume} {84}},\ \bibinfo {pages} {958--968} (\bibinfo
  {year} {2016})}\BibitemShut {NoStop}%
\bibitem [{\citenamefont {Samarakoon}\ \emph {et~al.}(2016)\citenamefont
  {Samarakoon}, \citenamefont {Sato}, \citenamefont {Chen}, \citenamefont
  {Chern}, \citenamefont {Yang}, \citenamefont {Klich}, \citenamefont
  {Sinclair}, \citenamefont {Zhou},\ and\ \citenamefont
  {Lee}}]{Samarakoon18102016}%
  \BibitemOpen
  \bibfield  {author} {\bibinfo {author} {\bibfnamefont {Anjana}\ \bibnamefont
  {Samarakoon}}, \bibinfo {author} {\bibfnamefont {Taku~J.}\ \bibnamefont
  {Sato}}, \bibinfo {author} {\bibfnamefont {Tianran}\ \bibnamefont {Chen}},
  \bibinfo {author} {\bibfnamefont {Gai-Wei}\ \bibnamefont {Chern}}, \bibinfo
  {author} {\bibfnamefont {Junjie}\ \bibnamefont {Yang}}, \bibinfo {author}
  {\bibfnamefont {Israel}\ \bibnamefont {Klich}}, \bibinfo {author}
  {\bibfnamefont {Ryan}\ \bibnamefont {Sinclair}}, \bibinfo {author}
  {\bibfnamefont {Haidong}\ \bibnamefont {Zhou}}, \ and\ \bibinfo {author}
  {\bibfnamefont {Seung-Hun}\ \bibnamefont {Lee}},\ }\bibfield  {title}
  {\enquote {\bibinfo {title} {Aging, memory, and nonhierarchical energy
  landscape of spin jam},}\ }\href {\doibase 10.1073/pnas.1608057113}
  {\bibfield  {journal} {\bibinfo  {journal} {Proceedings of the National
  Academy of Sciences}\ }\textbf {\bibinfo {volume} {113}},\ \bibinfo {pages}
  {11806--11810} (\bibinfo {year} {2016})},\ \Eprint
  {http://arxiv.org/abs/http://www.pnas.org/content/113/42/11806.full.pdf}
  {http://www.pnas.org/content/113/42/11806.full.pdf} \BibitemShut {NoStop}%
\end{thebibliography}%

\section{Appendix: High temperature expansion}
\label{sec:high_temp}
Here we derive \EQ{HighTemp} for the asymptotic $T_{b}^{-1}$ behavior of the probability of a direct strong Mpemba effect. The starting point is 
\bal &
\nonumber
\text{Prob}(\mathcal{P}(\mathcal{I}^{dir}_M)>0)
\\   
&=  \frac{1}{2}+\frac{1}{\pi }\text{sign}(\bm u^{dir}\cdot \bm w)\arctan \frac{1}{K},
\eal
where $K$ is given in \EQ{Kexpression}. By plugging in \EQS{udiv2}{wvec} in \EQ{Kexpression} we get
\bal
&
\nonumber
K
=
\\
&
\sqrt{\frac{\left(\sum_{i=1}^L e^{\beta_bE_i} \left(1-\frac{L}{Z(T_b)}e^{-\beta_bE_i} \right)^2 \right) \langle {\Delta E}^2 \rangle_b }{ \left(\sum_{j=1}^L \left(-E_j+\langle E\rangle_b \right)\right)^2}-1},\!
\eal
where $\langle E\rangle_b \equiv \sum _i \pi _i (T_b) E_i$ and $\langle \Delta E ^2 \rangle _b \equiv \sum _i \pi _i (T_b) (E_i - \langle E \rangle_b)^2$. At the high temperature limit, $T_b \to \infty$, we can expand $K$ in small $\beta_b$. To get the correct result, we have to expand all terms in the argument for the square root up to order $\beta_{b}^{2}$. Using 
$\arctan \frac{1}{K}\sim \frac{\pi}{2}-K,$
we find
\bal
{\rm Prob}[-(a_{2}( T=\infty )[\partial _{T}a_{2}]_{T=T_b} )>0]=\frac{C_{E}}{T_b},
\eal
where  
\bal  \nonumber
&C_{E}={\frac{1}{\pi} |(\bar{E}^2-\overline{E^2})
   |}\left(8 \bar{E}^6-24 \bar{E}^4 \overline{E^2}+20 \bar{E}^2
   \overline{E^2}^2-5 \overline{E^2}^3\right.  \\  &
   \left.+4 \bar{E}^3
   \overline{E^3}-  2 \bar{E} \overline{E^2}
   \overline{E^3} -\overline{E^3}^2-\bar{E}^2
   \overline{E^4}+\overline{E^2} \overline{E^4}\right)
   ^{1/2}
\eal
and $\overline{E^k}$ is the $k-$th moment of the energy distribution, defined as
\bal
\overline{E^k}\equiv {1\over L}\sum_{i=1}^{L}E_{i}^{k}.
\eal

\section{Appendix: Proof of Realizability of the Isotropic ensemble}
\label{sec:proof}

{\it Theorem}: Given any choice of a real vector $\bm f_{2}$ orthogonal to $\bm f_{1}$ in \eqref{eq:fone}, there exists a set of barriers $B_{ij}$ with relaxation dynamics obeying detailed balance \eqref{eq:Driving} having $F^{-1/2}\bm f_{2}$ as its slowest relaxation eigenvector.

{\it Proof}:
For our purpose we need to demonstrate at least one choice of barriers. We first note that for any (symmetrized) form of the driving $\tilde{R}$ with a steady state distribution $\bm f_{1}$, we can obtain, using \eqref{eq:Driving} and \eqref{eq:SymDriving}, formally, a set of barriers as:
\bal
B_{ij}=-\frac{1}{\beta_b}\Big(\log(\tilde{R}_{ij})-\frac{E_i+E_j}{2}\Big),~~~~i\neq j.\label{eq:barrierFromDrive}
\eal
The only requirement for these $B_{ij}$ to be consistent with our relaxation dynamics is that $B_{ij}$ is a real and symmetric matrix. In other words, it is sufficient that $\tilde{R}_{ij}$ is symmetric, and that $\tilde{R}_{ij}>0$ for all $i\neq j$ (note that $R_{ii}$ is then uniquely determined by the condition that $\bm f_{1}$ is an eigenvector with eigenvalue $0$).

We now show that we can make such a choice for any $\bm f_{2}$. To do so we consider first an initial set of barriers $B_{ij}= E_i+E_j$. An explicit calculation shows that the resulting dynamics has a single zero eigenvalue associated with $\bm f_{1}$, and that the rest of the eigenvalues are all $-Z(T_b)$. In this case we have $\tilde{R}_{ij}=e^{-\frac{\beta_b(E_i+E_j)}{2}}$. In particular any choice of $\bm f_{2}$ orthogonal with $\bm f_{1}$ is immediately an eigenvector of $\tilde{R}$. It remains to break the degeneracy between $\bm f_{2}$ and the other vectors orthogonal to $\bm f_{1}$. We do this  by adding a small perturbation to $\tilde{R}$:
\bal
\tilde{R}_{ij}\rightarrow \tilde{R}_{ij}+\frac{\epsilon}{||{\bm f_{2}}||^2} ({\bm f_{2}})_i ({\bm f_{2}})_j.
\eal
This change will only affect the eigenvalue associated with ${\bm f_{2}}$, changing it to $-Z(T_b)+\epsilon$, making it a non degenerate eigenvector.

Clearly, for $\epsilon$ small enough the positivity of $\tilde{R}_{ij}$ for $i\neq j$ will not be affected and the formula \eqref{eq:barrierFromDrive} will give us a valid set of barriers. (It is enough to take $\epsilon <{\rm min}_{ij}(e^{-\frac{\beta_b(E_i+E_j)}{2}})$). {\it QED}.

Of course, the above procedure yield a very particular type of barriers for each ${\bm f_{2}}$. There are numerous ways to set up other barriers consistent with a given ${\bm f_{2}}$.
\end{document}